\documentclass[journal,10pt]{IEEEtran}
\usepackage{amssymb}
\usepackage{amsmath}
\usepackage{cite}
\usepackage{url}
\usepackage{xcolor}
\usepackage{cite,graphicx,amsmath,amssymb}
\usepackage{subfigure}
\usepackage{citesort}
\usepackage{fancyhdr}
\usepackage{mdwmath}
\usepackage{mdwtab}
\usepackage{caption}
\usepackage{amsthm}
\usepackage{algorithm}
\usepackage{algorithmic}

\newtheorem{remark}{Remark}
\newtheorem{theorem}{Theorem}

\newtheorem{lemma}{Lemma}

\newtheorem{corollary}{Corollary}

\newtheorem{proposition}{Proposition}
\allowdisplaybreaks
\makeatletter
\def\ScaleIfNeeded{%
\ifdim\Gin@nat@width>\linewidth \linewidth \else \Gin@nat@width
\fi } \makeatother

\begin{document}

\title{Energy-Constrained UAV Data Collection Systems: NOMA and OMA}

\author{

Xidong~Mu,~\IEEEmembership{Graduate Student Member,~IEEE,}
        Yuanwei~Liu,~\IEEEmembership{Senior Member,~IEEE,}
        Li~Guo,~\IEEEmembership{Member,~IEEE,}
        Jiaru~Lin,~\IEEEmembership{Member,~IEEE,}
        and Zhiguo Ding,~\IEEEmembership{Fellow,~IEEE}

\thanks{Copyright (c) 2015 IEEE. Personal use of this material is permitted. However, permission to use this material for any other purposes must be obtained from the IEEE by sending a request to pubs-permissions@ieee.org.}
\thanks{X. Mu, L. Guo, and J. Lin are with the Key Laboratory of Universal Wireless Communications, Ministry of Education, Beijing University of Posts and Telecommunications, Beijing 100876, China, and are also with the School of Artificial Intelligence, Beijing University of Posts and Telecommunications, Beijing 100876, China. (email:\{muxidong, guoli, jrlin\}@bupt.edu.cn).}
\thanks{Y. Liu is with the School of Electronic Engineering and Computer Science, Queen Mary University of London, London, UK. (email:yuanwei.liu@qmul.ac.uk).}
\thanks{Z. Ding is with the School of Electrical and Electronic Engineering, The University of Manchester, Manchester, UK (e-mail:
zhiguo.ding@manchester.ac.uk).}
}

\maketitle
\begin{abstract}
This paper investigates unmanned aerial vehicle (UAV) data collection systems with different multiple access schemes, where a rotary-wing UAV is dispatched to collect data from multiple ground nodes (GNs). Our goal is to maximize the minimum UAV data collection throughput from GNs for both orthogonal multiple access (OMA) and non-orthogonal multiple access (NOMA) transmission, subject to the energy budgets at both the UAV and GNs, namely \emph{double energy limitations}. 1) For OMA, we propose an efficient algorithm by invoking alternating optimization (AO) method, where each subproblem is alternately solved by applying successive convex approximation (SCA) technique. 2) For NOMA, we first handle subproblems with fixed decoding order using SCA technique. Then, we develop a penalty-based algorithm to solve the decoding order design subproblem. Numerical results show that: i) The proposed algorithms are capable of improving the max-min throughput performance compared with other benchmark schemes; and ii) NOMA yields a higher performance gain than OMA when GNs have sufficient energy.
\end{abstract}
\section{Introduction}
Recently, unmanned aerial vehicles (UAVs) or drones have received extensive attention in military and civil applications with the advantages of low cost, high maneuverability, and high mobility~\cite{Wang_mag}. Equipped with communication devices, UAVs can act as aerial base stations (BSs) or users for accomplishing variant tasks, such as communication enhancement and offloading, cargo delivery, and security surveillance~\cite{Zeng2016Wireless,Mozaffari2019Tutorial,Zeng_Tutorial}. Among others, UAV data collection has been envisioned as a promising application, where UAVs equipped with sensing devices are deployed to collect data from ground nodes (GNs)~\cite{Motlagh2017}. Compared with terrestrial data collection systems, on the one hand, GNs are able to upload data to UAVs directly through the line-of-sight (LoS) dominated UAV-ground channels~\cite{Hourani2014}, thus reducing energy consumptions of GNs and improving the transmission efficiency. One the other hand, thanks to the low cost and high flexibility features of UAVs, UAV data collection systems are more suitable to be used in inaccessible regions, such as forest or marine monitoring, where deploying conventional terrestrial infrastructure to collect data is costly and inefficient.

Multiple access (MA) technique is one of the most fundamental enablers for facilitating UAV data collection systems since the number of GNs is usually large. The existing MA techniques can be loosely classified into two categories, namely, orthogonal multiple access (OMA) and non-orthogonal multiple access (NOMA). Different from OMA where one resource block (in time, frequency, or code) is occupied with at most one user, the key idea of NOMA\footnote{In this article, we use ``NOMA'' to refer to ``power-domain NOMA'' for simplicity.} is to allow different users to share the same time/frequency resources and to be multiplexed in different power levels by invoking superposition coding and successive interference cancellation (SIC) techniques~\cite{Liu2017,Cai_survey}. Owing to flexible resource allocations, NOMA is capable of improving spectrum efficiency, supporting massive connectivity, and guaranteing user fairness~\cite{Liu2017}. Recall that the diversified communication demands and massive connectivity requirements of UAV data collection systems, it is natural to investigate the employment of NOMA in UAV data collection systems and explore the potential performance gain.
\subsection{Related Works}
\subsubsection{Studies on UAV Communication Systems}
UAV communication systems have drawn significant attention of researchers in the past few years. In existing literature, UAVs are deployed as aerial BSs, relays, and users to boost the performance of communication systems, such as coverage and capacity. In terms of UAVs' state in the sky, research contributions can be divided into two categories: static UAV and mobile UAV communication systems. For static UAV communication systems, researchers mainly focused on the optimal deployment/placement of UAVs due to the unique air-to-ground (A2G) channel characteristics. The authors of~\cite{Hourani2014Optimal} studied the optimal UAV altitude to maximize the coverage based on the probability A2G channel model. The authors of~\cite{Mozaffari2016Efficient} further investigated optimal three-dimensional (3D) deployment of multiple UAVs, where an efficient method was proposed to achieve the maximum coverage while considering the inter-cell interference caused by different UAVs. The authors of~\cite{Lyu2017Placement} proposed a spiral-based algorithm with the aime of using the minimum number of UAVs to ensure that all ground users can be served. UAVs in coexistence with device-to-device (D2D) communications were studied by the authors of~\cite{Mozaffari2016Unmanned}, where the user outage probability was analyzed in both static and mobile UAV scenarios. For mobile UAV communication systems, the mobility of UAVs was exploited to further improve the system performance, such as average throughput and secrecy rate. A multiple UAV BSs network was considered by the authors of~\cite{Wu2018}, where the minimum average rate of ground users were maximized by optimizing UAVs' trajectories, transmit power, and user scheduling. The authors of~\cite{Cai2018} maximized the system secrecy rate by designing UAVs' trajectory and scheduling, where two UAVs were used for information transmission and jamming, respectively. The authors of~\cite{You2019} optimized 3D UAV trajectory in UAV-enabled data harvesting system with Rician fading channel model. Furthermore, a propulsion energy consumption model for rotary-wing UAVs were derived by the authors of~\cite{Zeng2019Energy}, where a novel path discretization method was proposed to minimize the energy consumed by the UAV for accomplishing missions. The authors of~\cite{Gong2018} studied a UAV flight time minimization problem in UAV data collection wireless sensor networks. The authors of~\cite{Zhan2019} minimized the energy consumption of Internet-of-Things (IoT) devices in a UAV-enabled IoT network, subject to the UAV energy constraint. The authors of~\cite{Sun2019Optimal} further studied solar-powered UAV communication systems, where the optimal UAV trajectory and resource allocations were obatined via monotonic optimization.
\subsubsection{Studies on UAV-NOMA Systems}
In contrast to the aforementioned research contributions on UAV communication systems considering OMA transmission scheme, some initial studies have focused on UAV-NOMA systems~\cite{Liu2019,Sohail2018,Hou2019,Mei2019,Duan,Cui2018Joint,Zhao2019,Mu_UAV}. For example, The authors of~\cite{Sohail2018} optimized the UAV attitude as well as power allocation to achieve maximum sum rate when UAV BSs serve ground users employing NOMA. The authors of~\cite{Hou2019} investigated multiple antennas technique in UAV NOMA communications, where the system performance was analyzed with stochastic geometry approach in both LOS and non-LoS scenarios. The authors of~\cite{Mei2019} proposed a novel uplink cooperative NOMA framework to tackle the interference introduced by the UAV user. A resource allocation problem was formulated by the authors of~\cite{Duan} in uplink NOMA multi-UAV IoT systems, where the system sum rate was maximized by optimizing subchannel allocation, IoT devices' transmit power, and UAVs' attitude. Furthermore, The authors of~\cite{Cui2018Joint} maximized the minimum achievable rate of ground users by jointly optimizing the UAV trajectory, transmit power, and user association in the downlink NOMA scenario. A UAV-assisted NOMA network was proposed by The authors of~\cite{Zhao2019}, where the UAV trajectory and precoding of the ground base station were jointly designed to maximize the system sum rate. The authors of~\cite{Mu_UAV} investigated uplink NOMA with the cellular-connected UAV, where the mission completion time was minimized by designing the UAV trajectory and UAV-BS association order. Moreover, the authors of \cite{Zhao_Security} developed two secure transmission schemes in UAV-NOMA networks for single-user and multiple-user scenarios. The authors of \cite{Zhao_mmWave} maximized the energy efficiency of mmWave-enabled NOMA-UAV networks by jointly optimizing the deployment location, hybrid precoding, and power allocation at the UAV.
\subsection{Motivation and Contributions}
Despite the aforementioned advantages of UAV data collection systems, one critical issue is that both the on-board energy of UAVs and storage energy of GNs are limited, namely double energy limitations. Therefore, the double energy limitations need to be carefully considered in practical designs to fully reap the benefits of UAV data collection systems. Although some prior works have studied the UAV data collection design \cite{You2019,Zeng2019Energy,Gong2018,Zhan2019}, the energy constraints at either UAV or GNs are often absent and only OMA transmission scheme was employed. To the best of our knowledge, the joint UAV trajectory and resource allocation design under different MA schemes has not been well investigated in the energy-constrained UAV data collection system, especially for NOMA. In contrast to OMA allocating orthogonal resources to different GNs, NOMA introduces additional decoding order design by multiplexing GNs in the same time/frequency resources, which makes the achievable data collection rate from GNs to the UAV more complicated and leads to a more challenging problem than OMA.

Against the above discussion, in this article, we investigate energy-constrained UAV data collection systems with two MA schemes, namely OMA and NOMA. Specifically, the UAV flies from the predefined initial location to the final location to harvest data from GNs, under the constraints on the UAV's and GNs' energy limitations. The main contributions are summarized as follows:
\begin{itemize}
  \item We propose a energy-constrained UAV data collection framework where both OMA and NOMA are employed at the UAV when collecting data from GNs. Based on the proposed framework, we jointly optimize the UAV trajectory, the GNs' transmit power, and the GN scheduling for maximization of the minimum UAV data collection throughput from all GNs, subject to the energy constraints at both the UAV and GNs.
  \item For OMA, we develop an efficient algorithm by employing alternating optimization (AO) method, where each non-convex subproblem is iteratively solved by applying successive convex approximation (SCA) technique. We demonstrate that the proposed algorithm is guaranteed to converge.
  \item For NOMA, we propose a penalty-based algorithm for solving the additional mixed integer non-convex decoding order design subproblem, where the relaxed continuous variables are forced to be binaries through iterations.
  \item Numerical results demonstrate that the max-min throughput obtained by the proposed algorithm significantly outperforms other benchmark schemes. It also shows that NOMA always achieves no worse performance than that of OMA, and the performance gain of NOMA over OMA is noticeable when GNs have sufficient energy.
\end{itemize}
\subsection{Organization and Notation}
The rest of the paper is organized as follows. Section II presents the system model and problem formulation for both OMA and NOMA. In Section III and Section IV, two efficient AO-based algorithms are developed for OMA and NOMA, respectively. Section V provides numerical results to validate the effectiveness of the proposed designs. Finally, Section VI concludes the paper.

\emph{Notation:} Scalars are denoted by lower-case letters, vectors are denoted by bold-face lower-case letters. ${\mathbb{R}^{M \times 1}}$ denotes the space of $M$-dimensional real-valued vector. For a vector ${\bf{a}}$, ${{\mathbf{a}}^T}$ denotes its transpose, and $\left\| {\mathbf{a}} \right\|$ denotes its Euclidean norm.
\section{System Model and Problem Formulation}
\subsection{System Model}
\begin{figure}[h!]
    \begin{center}
        \includegraphics[width=2.5in]{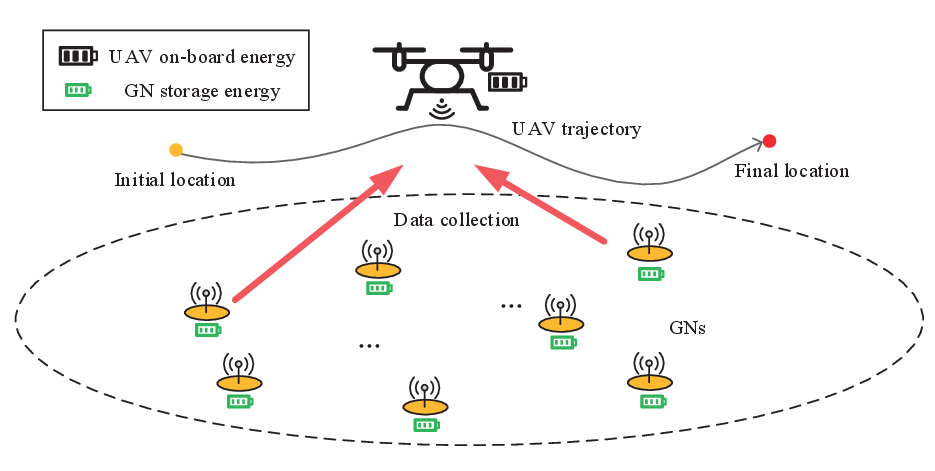}
        \caption{Illustration of the energy-constrained UAV data collection system.}
        \label{System model}
    \end{center}
\end{figure}
As shown in Fig. \ref{System model}, we consider a UAV data collection system, which consists of a rotary-wing single-antenna UAV data collector and $K$ single-antenna GNs. The GNs are indexed by the set ${\mathcal{K}} = \left\{ {1, \cdots ,K} \right\}$. The UAV is dispatched to fly from the predefined initial location to the final location with a constant height $H$. During the flight, the UAV collects data from each GN. Without loss of generality, a 3D Cartesian coordinate system is considered. The $k$th GN is fixed at ${\left( {{\mathbf{w}}_k^T,0} \right)^T}$, where ${{\mathbf{w}}_k} = {\left( {{x_k},{y_k}} \right)^T}$ denotes the corresponding horizontal coordinate. Similarly, $\left( {{\mathbf{u}}_I^T,H} \right)^T$ and $\left( {{\mathbf{u}}_F^T,H} \right)^T$ are the UAV's predefined initial and finial coordinates, where ${{\mathbf{u}}_I} = {\left( {{x_I},{y_I}} \right)^T}$ and ${{\mathbf{u}}_F} = {\left( {{x_F},{y_F}} \right)^T}$. Let $E_U$ and $E_k$ denoted the UAV total on-board energy and the total storage energy of the $k$th GN, respectively. Let $T_U$ denote the corresponding UAV total flight time, the instant UAV trajectory is denoted by ${\left( {{\mathbf{u}}{{\left( t \right)}^T},H} \right)^T},0 \le t \le {T_U}$, where ${\mathbf{u}}\left( t \right) \in {{\mathbb{R}}^{2 \times 1}}$ is the UAV horizontal location. Different from the existing UAV trajectory design works using the time discretization method~\cite{Wu2018,Cai2018,You2019}, $T_U$ is unknown and needs to be optimized in our work. To facilitate the design of UAV trajectory, we invoke the path discretization method~\cite{Zeng2019Energy} and divide the UAV path into $N$ line segments with $N+1$ waypoints, where ${\mathbf{u}}\left[ 1 \right] = {{\mathbf{u}}_I}$, ${\mathbf{u}}\left[ N+1 \right] = {{\mathbf{u}}_F}$. In order to achieve good approximation, we have the following constraints:
\begin{align}\label{path discretization}
\left\| {{\mathbf{u}}\left[ {n + 1} \right] - {\mathbf{u}}\left[ n \right]} \right\| \le \delta ,n = 1, \cdots ,N,
\end{align}
where $\delta $ is chosen sufficiently small compared with the UAV height such that the distance between the UAV and each GN is approximately unchanged and the UAV's speed can be regard as a constant within each line segment. The time duration of $n$th line segment is denoted by ${T\left[ n \right]}$ and $\sum\nolimits_{n = 1}^N {T\left[ n \right] = {T_U}} $. The constraints introduced by the UAV mobility is given by
\begin{align}\label{UAV Velocity Constraint}
\left\| {{{\mathbf{u}}}\left[ {n + 1} \right] - {{\mathbf{u}}}\left[ n \right]} \right\| \le {V_{\max }}{T\left[ n \right]} ,n = 1, \cdots ,N,
\end{align}
where ${V_{\max }}$ denotes the maximum speed of the UAV.

The channel coefficient between the UAV and the $k$th devices at the $n$th line segment can be modeled as ${h_k}\left[ n \right] = \sqrt {{\rho _k}\left[ n \right]} {\widetilde h_k}\left[ n \right]$, where ${{\rho _k}\left[ n \right]}$ represents the distance-dependent large-scale channel attenuation and ${\widetilde h_k}\left[ n \right]$ represents the small-scale fading coefficient. Recent A2G channel modeling literatures~\cite{3GPP_UAV,Matolak} have shown that there is a high probability for A2G channel to be dominated by the LoS link, especially for rural or suburban environment\footnote{As reported in~\cite{3GPP_UAV}, a 100\% LoS probability A2G channel can be achieved when the height of UAV is larger than 40 m in the rural macro (RMa) scenario.}, which is also the typical scenario for UAV data collection systems. Therefore, in this paper, we assume that ${\widetilde h_k}\left[ n \right] \triangleq 1$\footnote{In practice, the A2G channel might involve some random parameters, which make the associated optimization problems are complicated to solve. The results in this paper with the deterministic LoS channel model serve as a valuable performance indicator for the considered system, and provide guidelines for practical implementation.} and the channel coefficient follows from the free-space path loss model, which can be expressed as
\begin{align}\label{Channel gains between UAV and S k}
{\left| {h_k^{}\left[ n \right]} \right|^2} = \frac{{{\rho _0}}}{{{{\left\| {{\mathbf{u}}\left[ n \right] - {{\mathbf{w}}_k}} \right\|}^2} + {H^2}}},
\end{align}
where ${{\rho _0}}$ is the channel power gain at the reference distance of 1 meter. It is also assumed that the Doppler effect caused by the UAV mobility is perfectly compensated at the receivers~\cite{Synchronization}.

The energy consumption of UAV involves two parts: the communication related energy and the propulsion energy. Compared with the propulsion energy, the communication related energy is much smaller, and is thus ignored in this paper. Furthermore, we adopt the propulsion power consumption model of rotary-wing UAVs in~\cite{Zeng2019Energy}, which is modeled as a function of the UAV's speed and ignores the UAV acceleration/deceleration energy consumption. This model is reasonable when the acceleration/deceleration duration only takes a small portion of the total UAV flight time. Suppose that a rotary-wing UAV flying at the speed of $V$, the corresponding propulsion power consumption can be calculated as~\cite{Zeng2019Energy}
\begin{align}\label{UAV Power Model}
  P\left( V \right) \!\!=\!\! {P_0}\!\left(\! {1 \!\!+\!\! \frac{{3{V^2}}}{{U_{tip}^2}}} \!\right) \!\!+\!\!{P_i}\!{\left(\! {\sqrt {1 \!\!+\!\! \frac{{{V^4}}}{{4v_0^4}}}\!\!  -\!\! \frac{{{V^2}}}{{2v_0^2}}} \!\right)^{1/2}}\!\! \!\!+\!\! \frac{1}{2}{d_0}\rho sA{V^3},
\end{align}
where ${P_0}$ and ${P_i}$ are two constants representing the blade profile power and induced power in hovering status, respectively. ${U_{tip}}$ represents the tip speed of the rotor blade and $v_0$ represents the mean rotor induced velocity in hover. ${d_0}$ and $s$ are the fuselage drag ratio and rotor solidity, respectively. In addition, $\rho$ and $A$ denote the air density and rotor disc area, respectively. During the $n$th line segment, the UAV speed can be calculated as $V\left[ n \right] = \frac{{s\left[ n \right]}}{{T\left[ n \right]}}$, where $s\left[ n \right] = \left\| {{\mathbf{u}}\left[ {n + 1} \right] - {\mathbf{u}}\left[ n \right]} \right\|$. Thus, the UAV energy consumption at the $n$th line segment is given by
\begin{align}\label{energy consumption}
\begin{gathered}
  E\left[ n \right] = T\left[ n \right]P\left( {\frac{{s\left[ n \right]}}{{T\left[ n \right]}}} \right) \hfill \\
  = {P_0}\left( {T\left[ n \right] + \frac{{3s{{\left[ n \right]}^2}}}{{U_{tip}^2T\left[ n \right]}}} \right) \hfill \\
  + {P_i}{\left( {\sqrt {T{{\left[ n \right]}^4} + \frac{{s{{\left[ n \right]}^4}}}{{4v_0^4}}}  - \frac{{s{{\left[ n \right]}^2}}}{{2v_0^2}}} \right)^{\frac{1}{2}}} + \frac{1}{2}{d_0}\rho sA\frac{{s{{\left[ n \right]}^3}}}{{T{{\left[ n \right]}^2}}}. \hfill \\
\end{gathered}
\end{align}
Then, the total UAV energy constraint can be expressed as $\sum\nolimits_{n = 1}^N {E\left[ n \right]}  \le {E_U}$.

The sleep-wake protocol is considered when the UAV collects data from GNs, as assumed in \cite{You2019,Zhan2019}. GNs upload information only when being waken up by the UAV, otherwise they keep in silence. In this paper, an offline design is considered, i.e., first determining the UAV trajectory and the wake-up time allocation scheme using the prior knowledge of GNs' locations and their LoS channel coefficients. Based on the obtained results, during the realistic flight, the UAV wakes up the corresponding GN\footnote{In terms of the wake-up protocol~\cite{Ratasuk}, GNs remain in power save state instead of completely off state. As a result, the start time is relatively short and can be ignored as compared with the data uploading duration.} at the predefined time instant and location using the wake-up signal~\cite{Ratasuk} and informs them to upload data through the downlink reliable control links. We ignore the energy consumed by the UAV for sending wake-up signal, since it belongs to the communication related energy. In the next subsection, we formulate the optimization problem with two MA schemes, i.e., OMA and NOMA.
\subsection{Problem Formulation for OMA}
For OMA, the UAV receives the information bits from different GNs by allocating unique time resources\footnote{We adopt time division multiple access (TDMA) instead of frequency division multiple access (FDMA) for OMA, since TDMA is more energy-saving and easier for implementation than FDMA in the considered network of this paper.} at each time duration $T\left[ n \right]$. Let ${\tau _k}\left[ n \right]$ denote the allocated time resources for the $k$th GN during $T\left[ n \right]$ and we have $\sum\nolimits_{k = 1}^K {{\tau _k}\left[ n \right]}  \le T\left[ n \right],\forall n$. Recall that the distance between the UAV and each GN is approximately unchanged within each line segment by employing the path discretization method. Therefore, the achievable data collection throughput (bits/Hz) from the $k$th GN during the $n$th line segment for OMA is approximate to a constant, which is given by
\begin{align}\label{throughput_OMA}
{r_k^{{\rm{O}}}}\left[ n \right] = {\tau _k}\left[ n \right]\log_2 \left( {1 + \frac{{{\gamma _0}{p_k}\left[ n \right]}}{{{{\left\| {{\mathbf{u}}\left[ n \right] - {{\mathbf{w}}_k}} \right\|}^2+ {H^2}}}}} \right),
\end{align}
where ${\gamma _0} = \frac{{{\rho _0}}}{{{\sigma ^2}}}$ and ${p_k}\left[ n \right]$ denotes the transmit power of the $k$th GN. Therefore, the total achievable throughput from the $k$th GN during the UAV flight is $Q_k^{{\rm{O}}} = \sum\nolimits_{n = 1}^N {r_k^{{\rm{O}}}\left[ n \right]} $. In our work, the circuit power consumptions of GNs are ignored, the total storage energy constraint of the $k$th GN for OMA is given by $\sum\nolimits_{n = 1}^N {{\tau _k}\left[ n \right]{p_k}\left[ n \right]}  \le {E_k},\forall k$.

In order to achieve a fair data collection from all GNs, we aim to maximize the minimum UAV data collection throughput by jointly optimizing the UAV trajectory, $\left\{ {{\mathbf{u}}\left[ n \right],T\left[ n \right]} \right\}$, the GN scheduling, $\left\{ {{\tau _k}\left[ n \right]} \right\}$, and the GN transmit power, $\left\{ {{p_k}\left[ n \right]} \right\}$, while taking the energy constraints of both the UAV and GNs into account. Then, the optimization problem for OMA can be formulated as
\begin{subequations}\label{P1}
\begin{align}
&\mathop {\max }\limits_{\left\{ {{\mathbf{u}}\left[ n \right],T\left[ n \right],{\tau _k}\left[ n \right],{p_k}\left[ n \right]} \right\}} \;\;\;\mathop {\min }\limits_{\forall k} Q_k^{{\rm{O}}} \\
\label{One UAV Initial Final Location Constraint OMA}{\rm{s.t.}}\;\;&{{\mathbf{u}}}\left[ 1 \right] = {{\mathbf{u}}_{I}},{{\mathbf{u}}}\left[ N \right] = {{\mathbf{u}}_{F}},\\
\label{One UAV Velocity Constraint OMA}&\left\| {{\mathbf{u}}\left[ {n + 1} \right] - {\mathbf{u}}\left[ n \right]} \right\| \le \min \left( {\delta ,{V_{\max }}T\left[ n \right]} \right),\forall n,\\
\label{UAV energy OMA}&\sum\nolimits_{n = {1}}^{{N}} {E\left[ n \right]}  \le {E_U},\\
\label{sensor energy OMA}&\sum\nolimits_{n = 1}^N {\tau _k \left[ n \right] {p_k}\left[ n \right]}  \le {E_k},\forall k,\\
\label{time duration OMA}&\sum\nolimits_{k = 1}^K {{\tau _k}\left[ n \right]}  \le T\left[ n \right],\forall n,\\
\label{time duration OMA 2}&{\tau _k}\left[ n \right] \ge 0,\forall k,n,\\
\label{transmit power OMA}&0 \le {p_k}\left[ n \right] \le {P_{\max }},\forall n,k,
\end{align}
\end{subequations}
where ${P_{\max }}$ represents the maximum transmit power of GNs. \eqref{One UAV Initial Final Location Constraint OMA} and \eqref{One UAV Velocity Constraint OMA} represent the UAV mobility constraints. \eqref{UAV energy OMA} and \eqref{sensor energy OMA} are energy constraints of the UAV and GNs, respectively.
\begin{remark}\label{remark:TDMA mode}
\emph{In Problem \eqref{P1}, time resources are allocated in an adaptive manner. We refer to this type of OMA as OMA-II scheme. However, in conventional OMA, time resources are equally allocated to each GN, which is referred to OMA-I scheme~\cite{Chen2017Optimization}. For OMA-I, Problem \eqref{P1} needs to consider additional constraints, ${\tau _k}\left[ n \right] = {\tau _j}\left[ n \right],\forall k \ne j$.}
\end{remark}
\subsection{Problem Formulation for NOMA}
For NOMA, the UAV receives all GNs' signals through the same time resources. Different from the downlink NOMA communication \cite{Cui2018Joint}, where the SIC decoding order is determined by the channel gains. In uplink NOMA, the UAV can perform SIC in any arbitrary order since all received signals at the UAV are desired signals. Let ${\pi _n}\left( k \right)$ denote the decoding order of GN $k$ at the $n$th line segment. If ${\pi _n}\left( k \right) = i$, then GN $k$ is the $i$th signal to be decoded at the $n$th line segment. Therefore, a set of binary indicators ${\alpha _{k,m}}\left[ n \right] \in \left\{ {0,1} \right\},\forall k \ne m \in {\mathcal{K}}$ are defined as
\begin{align}\label{decoding order1}
&{\alpha _{k,m}}\left[ n \right] = 1,{\pi _n}\left( k \right) > {\pi _n}\left( m \right),\\
\label{decoding order2}&{\alpha _{k,m}}\left[ n \right] + {\alpha _{m,k}}\left[ n \right] = 1.
\end{align}
Equation \eqref{decoding order2} ensures that there is only one GN at each decoding order.

Similarly, the UAV received signal-to-interference-plus-noise (SINR) from the $k$th GN during the $n$th line segment can be approximate to a constant, which is given by
\begin{align}\label{SINR NOMA}
\begin{gathered}
  {\gamma _k^{{\rm{N}}}}\left[ n \right] = \frac{{{\gamma _0}{p_k}\left[ n \right]{{\left| {{h_k}\left[ n \right]} \right|}^2}}}{{\sum\nolimits_{m \in {{\mathcal{K}}},m \ne k} {{\alpha _{m,k}}\left[ n \right]{\gamma _0}{p_m}\left[ n \right]{{\left| {{h_m}\left[ n \right]} \right|}^2} + 1} }} \hfill \\
  = \!\frac{{{{{\gamma _0}{p_k}\left[ n \right]} \mathord{\left/
 {\vphantom {{{\gamma _0}{p_k}\left[ n \right]} {\left( {{{\left\| {{\mathbf{u}}\left[ n \right]\!\! -\!\! {{\mathbf{w}}_k}} \right\|}^2} \!+\! {H^2}} \right)}}} \right.
 \kern-\nulldelimiterspace} {\left( {{{\left\| {{\mathbf{u}}\left[ n \right]\!\! -\!\! {{\mathbf{w}}_k}} \right\|}^2} \!+\! {H^2}} \right)}}}}{{\sum\nolimits_{m \in {{\mathcal{K}}},m \ne k} {{{{\alpha _{m,k}}\left[ n \right]{\gamma _0}{p_m}\left[ n \right]} \mathord{\left/
 {\vphantom {{{\alpha _{m,k}}\left[ n \right]{\gamma _0}{p_m}\left[ n \right]} {\left( {{{\left\| {{\mathbf{u}}\left[ n \right]\!\! -\!\! {{\mathbf{w}}_m}} \right\|}^2} \!+\! {H^2}} \right)}}} \right.
 \kern-\nulldelimiterspace} {\left( {{{\left\| {{\mathbf{u}}\left[ n \right]\!\! -\! \!{{\mathbf{w}}_m}} \right\|}^2} \!+\! {H^2}} \right)}} \!+\! 1} }}. \hfill \\
\end{gathered}
\end{align}
Let $\tau \left[ n \right]$ denote the UAV allocated time resources for all GNs at the $n$th line segment, where $\tau \left[ n \right] \le T\left[ n \right],\forall n$. Similarly, the total data collection throughput from the $k$th GN during the UAV flight for NOMA is given by
\begin{align}\label{achievable rate UAV k}
  {Q_k^{{\rm{N}}}} \!=\! \sum\nolimits_{n = 1}^N {r_k^{{\rm{N}}}\left[ n \right]}\! =\! \sum\nolimits_{n = 1}^N {\tau \left[ n \right]{{\log }_2}\left( {1 \!+ \!{\gamma _k^{{\rm{N}}}}\left[ n \right]} \right)} .
\end{align}
Moreover, the GNs' energy constraints for NOMA are given by $\sum\nolimits_{n = 1}^N {\tau \left[ n \right]{p_k}\left[ n \right]}  \le {E_k},\forall k.$

By jointly optimizing the UAV trajectory, $\left\{ {{\mathbf{u}}\left[ n \right],T\left[ n \right]} \right\}$, the communication time allocation, $\left\{ {{\tau }\left[ n \right]} \right\}$, the GN transmit power, $\left\{ {{p_k}\left[ n \right]} \right\}$, and the decoding order, $\left\{ {{\alpha _{k,m}}\left[ n \right]} \right\}$, the max-min collection throughput optimization problem for NOMA can be formulated as
\begin{subequations}\label{P2}
\begin{align}
&\mathop {\max }\limits_{\left\{ {{\mathbf{u}}\left[ n \right],T\left[ n \right],\tau \left[ n \right],{p_k}\left[ n \right],{\alpha _{k,m}}\left[ n \right]} \right\}} \;\;\;\mathop {\min }\limits_{\forall k} Q_k^{{\rm{N}}} \\
\label{One UAV Initial Final Location Constraint NOMA}{\rm{s.t.}}\;\;&{{\mathbf{u}}}\left[ 1 \right] = {{\mathbf{u}}_{I}},{{\mathbf{u}}}\left[ N \right] = {{\mathbf{u}}_{F}},\\
\label{One UAV Velocity Constraint NOMA}&\left\| {{\mathbf{u}}\left[ {n + 1} \right] - {\mathbf{u}}\left[ n \right]} \right\| \le \min \left( {\delta ,{V_{\max }}T\left[ n \right]} \right),\forall n,\\
\label{UAV energy NOMA}&\sum\nolimits_{n = {1}}^{{N}} {E\left[ n \right]}  \le {E_U},\\
\label{sensor energy NOMA}&\sum\nolimits_{n = 1}^N {\tau \left[ n \right] {p_k}\left[ n \right]}  \le {E_k},\forall k,\\
\label{time duration NOMA}&0 \le {{\tau }\left[ n \right]}  \le T\left[ n \right],\forall n,\\
\label{transmit power NOMA}&0 \le {p_k}\left[ n \right] \le {P_{\max }},\forall n,k,\\
\label{decoding order NOMA}&{\alpha _{k,m}}\left[ n \right] + {\alpha _{m,k}}\left[ n \right] = 1,\forall k \ne m \in {{\mathcal{K}}},\\
\label{index NOMA}&{\alpha _{k,m}}\left[ n \right] \in \left\{ {0,1} \right\},\forall k, m \in \mathcal{K},
\end{align}
\end{subequations}

Problems \eqref{P1} and \eqref{P2} are non-convex problems due to the non-convex objective function and constraints, where optimization variables are highly-coupled. Moreover, the introduced binary variables for NOMA decoding orders make \eqref{P2} become a mixed integer non-convex optimization problem, which is more challenging to solve. Note that there is no standard method to efficiently obtain the globally optimal solution for such problems. In the following, we develop efficient algorithms to find a high-quality suboptimal solution with a polynomial time complexity, by employing AO method and SCA technique.
\section{Proposed Solution for OMA}
To make Problem \eqref{P1} tractable, we first introduce auxiliary variables $\left\{ {{\theta _k}\left[ n \right]},\forall n,k \right\}$ and $\left\{ {\omega \left[ n \right] \ge 0} ,\forall n\right\}$ such that
\begin{align}
\label{theta OMA}&{\theta _k}{\left[ n \right]^2} = {\tau _k}\left[ n \right]B{\log _2}\left( {1 + \frac{{{\gamma _0}{p_k}\left[ n \right]}}{{{{\left\| {{\mathbf{u}}\left[ n \right] - {{\mathbf{w}}_k}} \right\|}^2} + {H^2}}}} \right),\\
\label{w OMA}&\omega \left[ n \right] = {\left( {\sqrt {T{{\left[ n \right]}^4} + \frac{{s{{\left[ n \right]}^4}}}{{4v_0^4}}}  - \frac{{s{{\left[ n \right]}^2}}}{{2v_0^2}}} \right)^{\frac{1}{2}}}.
\end{align}
Moreover, equation \eqref{w OMA} is equivalent to
\begin{align}\label{w1}
\frac{{T{{\left[ n \right]}^4}}}{{\omega {{\left[ n \right]}^2}}} = \omega {\left[ n \right]^2} + \frac{{s{{\left[ n \right]}^2}}}{{v_0^2}}.
\end{align}
With the above introduced variables and define ${\eta ^{{\rm{O}}}} = \mathop {\min }\limits_{\forall k} Q_k^{{\rm{O}}}$, Problem \eqref{P1} can be rewritten as the following problem:
\begin{subequations}\label{P1.1}
\begin{align}
&\mathop {\max }\limits_{{\eta ^{{\rm{O}}}},\left\{ {{\mathbf{u}}\left[ n \right],T\left[ n \right],{\tau _k}\left[ n \right],{p_k}\left[ n \right],{\theta _k}\left[ n \right],\omega \left[ n \right]} \right\}} \;\;\;{\eta ^{{\rm{O}}}} \\
\label{throughput OMA P1.1}{\rm{s.t.}}\;\;&\sum\nolimits_{n = 1}^N {\theta _k}{\left[ n \right]^2}  \ge {\eta ^{{\rm{O}}}} ,\forall k,\\
\label{theta OMA P1.1}&\frac{{{\theta _k}{{\left[ n \right]}^2}}}{{{\tau _k}\left[ n \right]}} \le {\log _2}\left( {1 + \frac{{{\gamma _0}{p_k}\left[ n \right]}}{{{{\left\| {{\mathbf{u}}\left[ n \right] - {{\mathbf{w}}_k}} \right\|}^2} + {H^2}}}} \right),\forall n,k,\\
\label{UAV total energy OMA P1.1}&\begin{gathered}
  {P_0}\sum\nolimits_{n = 1}^N {\left( {T\left[ n \right] + \frac{{3s{{\left[ n \right]}^2}}}{{U_{tip}^2T\left[ n \right]}}} \right)}  + {P_i}\sum\nolimits_{n = 1}^N {\omega \left[ n \right]}  \hfill \\
  \;\;\;\;\;\;\;\;\;\;\;\;\;\;\;\;\; + \frac{1}{2}{d_0}\rho sA\sum\nolimits_{n = 1}^N {\frac{{s{{\left[ n \right]}^3}}}{{t{{\left[ n \right]}^2}}}}  \le {E_U}, \hfill \\
\end{gathered} \\
\label{w OMA P1.1}&\frac{{T{{\left[ n \right]}^4}}}{{\omega {{\left[ n \right]}^2}}} \le \omega {\left[ n \right]^2} + \frac{{{{\left\| {{\mathbf{u}}\left[ {n + 1} \right] - {\mathbf{u}}\left[ n \right]} \right\|}^2}}}{{v_0^2}},\forall n,\\
\label{constraints OMA P1.1}&\eqref{One UAV Initial Final Location Constraint OMA},\eqref{One UAV Velocity Constraint OMA},\eqref{sensor energy OMA}-\eqref{transmit power OMA}.
\end{align}
\end{subequations}
\begin{proposition}\label{optimal transmit power}
\emph{Problems \eqref{P1.1} and \eqref{P1} are equivalent.}
\begin{proof}
\emph{Without loss of optimality to Problem \eqref{P1.1}, constraints \eqref{theta OMA P1.1} and \eqref{w OMA P1.1} can be met with equality. Specifically, assume that if any of constraints in \eqref{theta OMA P1.1} is satisfied with strict inequality, then we can always increase the corresponding value of ${\theta _k}{\left[ n \right]^2}$ to make the constraint \eqref{theta OMA P1.1} satisfied with equality without decreasing the objective value. Furthermore, suppose that \eqref{w OMA P1.1} are satisfied with strict inequality, we can always reduce the corresponding value of $\omega\left[ n \right]$ to make the constraint \eqref{w OMA P1.1} satisfied with equality with other variables fixed, and at the same time make the constraint \eqref{UAV total energy OMA P1.1} still satisfied without changing the objective value of \eqref{P1.1}. Therefore, Problems \eqref{P1.1} and \eqref{P1} are equivalent.}
\end{proof}
\end{proposition}

Based on \textbf{Proposition 1}, we only need to focus on how to solve Problem \eqref{P1.1}. As the first and third terms in \eqref{UAV total energy OMA P1.1} are the perspective of the convex quadratic function and the convex cubic function on the set of real numbers, they are also joint convex functions with respect to $s\left[ n \right]$ and $T\left[ n \right]$ since the perspective operation preserves convexity~[36, Page 89]. As a result, \eqref{UAV total energy OMA P1.1} is a convex constraint. However, constraints \eqref{sensor energy OMA}, \eqref{throughput OMA P1.1}, \eqref{theta OMA P1.1}, and \eqref{w OMA P1.1} are still non-convex, where the optimization variables are high-coupled. To handle this obstacle, in the following, we propose AO-based algorithm. Specifically, the original problem \eqref{P1.1} is decomposed into two subproblems, i.e., optimization with fixed transmit power and optimization with fixed UAV locations, which are handled by employing SCA. The two subproblems are alternatingly solved until convergence.
\subsection{Optimization with Fixed Transmit Power}
For any given feasible GN transmit power, $\left\{ {{p_k}\left[ n \right]} \right\}$, the optimization problem can be written as
\begin{subequations}\label{P1.2}
\begin{align}
&\mathop {\max }\limits_{{\eta ^{{\rm{O}}}},\left\{ {{\mathbf{u}}\left[ n \right],T\left[ n \right],{\tau _k}\left[ n \right],{\theta _k}\left[ n \right],\omega \left[ n \right]} \right\}} \;\;\;{\eta ^{{\rm{O}}}}  \\
\label{constraints P1.2}{\rm{s.t.}}\;\;&\eqref{One UAV Initial Final Location Constraint OMA},\eqref{One UAV Velocity Constraint OMA},\eqref{sensor energy OMA}-\eqref{time duration OMA},\eqref{throughput OMA P1.1}-\eqref{w OMA P1.1}.
\end{align}
\end{subequations}
Problem \eqref{P1.2} is still a non-convex problem due to the non-convex constraints \eqref{throughput OMA P1.1}, \eqref{theta OMA P1.1}, and \eqref{w OMA P1.1}. Fortunately, those non-convex constraints can be handled by utilizing SCA technique. Specifically, to tackle the non-convex constraint \eqref{throughput OMA P1.1}, the left hand side (LHS) is a convex function with respect to ${\theta _k}\left[ n \right]$. Since any convex functions are lower bounded by their first-order Taylor expansion, the lower bound of ${\theta _k}{\left[ n \right]^2}$ at a given local point $\theta _k^l\left[ n \right]$ can be expressed as
\begin{align}\label{throughput OMA lower bound}
{\theta _k}{\left[ n \right]^2} \ge \theta _k^l{\left[ n \right]^2} + 2\theta _k^l\left[ n \right]\left( {\theta _k^{}\left[ n \right] - \theta _k^l\left[ n \right]} \right).
\end{align}

For the non-convex constraint \eqref{theta OMA P1.1}, the LHS is jointly convex with respect to ${{\theta _k}\left[ n \right]}$ and ${{\tau _k}\left[ n \right]}$. Though the right hand side (RHS) is not concave with respect to ${{\mathbf{u}}\left[ n \right]}$, it is a convex function with respect to ${{{\left\| {{\mathbf{u}}\left[ n \right] - {{\mathbf{w}}_k}} \right\|}^2}}$. With given local points $\left\{ {{{\mathbf{u}}^l}\left[ n \right]} \right\}$, the lower bound for the RHS of \eqref{theta OMA P1.1} can be expressed as
\begin{align}\label{tau Rk}
\begin{gathered}
  \log_2 \left( {1 + \frac{{{\gamma _0}p_k\left[ n \right]}}{{{{\left\| {{\mathbf{u}}\left[ n \right] - {{\mathbf{w}}_k}} \right\|}^2} + {H^2}}}} \right) \ge R_k^{lb}\left[ n \right] \hfill \\
   = \log_2 \left( {1 + \frac{{{\gamma _0}{p_k}\left[ n \right]}}{{{{\left\| {{{\mathbf{u}}^l}\left[ n \right] - {{\mathbf{w}}_k}} \right\|}^2} + {H^2}}}} \right) \hfill \\
  \;\;\;\;\;\;\;\;\;\; - \varphi _k^l\left[ n \right]\left( {{{\left\| {{\mathbf{u}}\left[ n \right] - {{\mathbf{w}}_k}} \right\|}^2} - {{\left\| {{{\mathbf{u}}^l}\left[ n \right] - {{\mathbf{w}}_k}} \right\|}^2}} \right), \hfill \\
\end{gathered}
\end{align}
where $\varphi _k^l\left[ n \right] = \frac{{\left( {{{\log }_2}e} \right){\gamma _0}{p_k}\left[ n \right]}}{{\left( {{{\left\| {{{\mathbf{u}}^l}\left[ n \right] - {{\mathbf{w}}_k}} \right\|}^2} + {H^2}} \right)\left( {{{\left\| {{{\mathbf{u}}^l}\left[ n \right] - {{\mathbf{w}}_k}} \right\|}^2} + {H^2} + {\gamma _0}{p_k}\left[ n \right]} \right)}}$.

Similarly, to deal with the non-convex constraint \eqref{w OMA P1.1}, the RHS is the sum of two convex functions and the lower bound is given by
\begin{align}\label{lower bound OMA}
\omega {\left[ n \right]^2} + \frac{{{{\left\| {{\mathbf{u}}\left[ {n + 1} \right] - {\mathbf{u}}\left[ n \right]} \right\|}^2}}}{{v_0^2}} \ge {\beta ^{lb}}\left[ n \right],
\end{align}
where
\[\begin{gathered}
  {\beta ^{lb}}\left[ n \right]\! =\! {\omega ^l}{\left[ n \right]^2} \!+\! {\omega ^l}\left[ n \right]\left( {\omega \left[ n \right] \!\!-\!\! {\omega ^l}\left[ n \right]} \right) \!-\! \frac{{{{\left\| {{{\mathbf{u}}^l}\left[ {n \!\!+\!\! 1} \right] \!\!-\!\! {{\mathbf{u}}^l}\left[ n \right]} \right\|}^2}}}{{v_0^2}} \hfill \\
  \;\;\;\;\;\;\;\;\;\;\;\;\;\;\;\;\;\;\;\;\;\;\;\;\;\; + \frac{2}{{v_0^2}}{\left( {{{\mathbf{u}}^l}\left[ {n \!\!+\!\! 1} \right] \!\!-\!\! {{\mathbf{u}}^l}\left[ n \right]} \right)^T}\left( {{\mathbf{u}}\left[ {n \!\!+\!\! 1} \right] \!\!-\!\! {\mathbf{u}}\left[ n \right]} \right). \hfill \\
\end{gathered} \]
By applying \eqref{throughput OMA lower bound}-\eqref{lower bound OMA}, Problem \eqref{P1.2} is approximated as the following optimization problem:
\begin{subequations}\label{P1.3}
\begin{align}
&\mathop {\max }\limits_{{\eta ^{{\rm{O}}}},\left\{ {{\mathbf{u}}\left[ n \right],T\left[ n \right],{\tau _k}\left[ n \right],{\theta _k}\left[ n \right],\omega \left[ n \right]} \right\}} \;\;\;{\eta ^{{\rm{O}}}}  \\
\label{SCA throughput OMA 1.3}{\rm{s.t.}}\;\;&\sum\nolimits_{n = 1}^N {\left( {\theta _k^l{{\left[ n \right]}^2} + 2\theta _k^l\left[ n \right]\left( {\theta _k^{}\left[ n \right] - \theta _k^l\left[ n \right]} \right)} \right)}  \ge \eta^{{\rm{O}}} ,\forall k,\\
\label{SCA theta OMA 1.3}&\frac{{{\theta _k}{{\left[ n \right]}^2}}}{{{\tau _k}\left[ n \right]}} \le R_k^{lb}\left[ n \right],\forall k,n,\\
\label{SCA w OMA 2}&\frac{{T{{\left[ n \right]}^4}}}{{\omega {{\left[ n \right]}^2}}} \le {\beta ^{lb}}\left[ n \right],\forall n,\\
\label{constraints OMA}&\eqref{One UAV Initial Final Location Constraint OMA},\eqref{One UAV Velocity Constraint OMA},\eqref{sensor energy OMA}-\eqref{time duration OMA},\eqref{UAV total energy OMA P1.1}.
\end{align}
\end{subequations}
Now, \eqref{SCA throughput OMA 1.3} is a linear constraint, and \eqref{SCA theta OMA 1.3} and \eqref{SCA w OMA 2} are all convex constraints. Therefore, Problem \eqref{P1.3} is a convex problem, which can be efficiently solved by standard convex optimization solvers such as CVX~\cite{cvx}. Due to the adoption of the lower bounds in \eqref{throughput OMA lower bound}-\eqref{lower bound OMA}, any feasible solution of Problem \eqref{P1.3} must be also feasible for Problem \eqref{P1.2}, but the reverse does not hold in general. This means the feasible set of problem \eqref{P1.3} is a smaller convex set reduced from the original non-convex feasible set of problem \eqref{P1.2}. Therefore, the obtained objective value of Problem \eqref{P1.3} in general provides a lower bound of that of Problem \eqref{P1.2}.
\subsection{Optimization with Fixed UAV Locations}
Before solving this subproblem, we introduce additional auxiliary variables $\left\{ {\varepsilon_k \left[ n \right]} ,\forall n,k\right\}$ such that
\begin{align}
\label{energy kn}&{\varepsilon _k}{\left[ n \right]^2} = {\tau _k}\left[ n \right]{p_k}\left[ n \right],\forall n,k.
\end{align}
Thus, constraint \eqref{sensor energy OMA} can be equivalently expressed as the following two constraints:
\begin{align}
\label{ep energy}&\sum\nolimits_{n = 1}^N {{\varepsilon _k}{{\left[ n \right]}^2}}  \le {E_k},\forall k,\\
\label{ep constraint}&{p_k}\left[ n \right] \le \frac{{{\varepsilon _k}{{\left[ n \right]}^2}}}{{{\tau _k}\left[ n \right]}},\forall n,k.
\end{align}
For any given feasible UAV location, ${\left\{ {{\mathbf{u}}\left[ n \right]} \right\}}$, Problem \eqref{P1.1} can be written as the following optimization problem:
\begin{subequations}\label{P1.4}
\begin{align}
&\mathop {\max }\limits_{\eta ^{{\rm{O}}},\left\{ {T\left[ n \right],{\tau _k}\left[ n \right],{p_k}\left[ n \right],{\theta _k}\left[ n \right],\omega \left[ n \right],{\varepsilon_k \left[ n \right]}} \right\}} \;\;\;{\eta ^{{\rm{O}}}} \\
\label{P1.4 constraints}{\rm{s.t.}}\;\;&\eqref{time duration OMA}-\eqref{transmit power OMA},\eqref{throughput OMA P1.1}-\eqref{w OMA P1.1},\eqref{ep energy},\eqref{ep constraint}.
\end{align}
\end{subequations}
Now, constraint \eqref{theta OMA P1.1} is convex since the RHS is a concave function with respect to ${{p_k}\left[ n \right]}$. However, Problem \eqref{P1.4} is still a non-convex problem due to the non-convex constraints \eqref{throughput OMA P1.1}, \eqref{w OMA P1.1}, and \eqref{ep constraint}. As described in the previous subsection, we have already introduced how to deal with non-convex constraints \eqref{throughput OMA P1.1} and \eqref{w OMA P1.1} with their lower bounds based on the first-order Taylor expansion. Therefore, we only need to concentrate on how to deal with the non-convex constraint \eqref{ep constraint}. Similarly, the RHS of \eqref{ep constraint} is jointly convex with respect to ${{\varepsilon _k}\left[ n \right]}$ and ${{\tau _k}\left[ n \right]}$. The lower bound with given local points $\left\{ {\varepsilon _k^l\left[ n \right],\tau _k^l\left[ n \right]} \right\}$ can be expressed as
\begin{align}\label{ep lower bound}
\begin{gathered}
  \frac{{{\varepsilon _k}{{\left[ n \right]}^2}}}{{{\tau _k}\left[ n \right]}} \ge \chi _k^{lb}\left[ n \right] = \frac{{\varepsilon _k^l{{\left[ n \right]}^2}}}{{\tau _k^l\left[ n \right]}} + \;\frac{{2\varepsilon _k^l\left[ n \right]}}{{\tau _k^l\left[ n \right]}}\left( {\varepsilon _k^{}\left[ n \right] - \varepsilon _k^l\left[ n \right]} \right) \hfill \\
  \;\;\;\;\;\;\;\;\;\;\;\;\;\;\;\;\;\;\;\;\;\;\;\;\;\;\;\; - \frac{{\varepsilon _k^l{{\left[ n \right]}^2}}}{{\tau _k^l{{\left[ n \right]}^2}}}\left( {\tau _k^{}\left[ n \right] - \tau _k^l\left[ n \right]} \right),\forall n,k, \hfill \\
\end{gathered}
\end{align}
By replacing these non-convex constraints with their lower bounds, Problem \eqref{P1.4} can be written as the following approximate optimization problem:
\begin{subequations}\label{P1.5}
\begin{align}
&\mathop {\max }\limits_{{\eta ^{{\rm{O}}}},\left\{ {T\left[ n \right],{\tau _k}\left[ n \right],{p_k}\left[ n \right],{\theta _k}\left[ n \right],\omega \left[ n \right],{\varepsilon_k \left[ n \right]}} \right\}} \;\;\;{\eta ^{{\rm{O}}}} \\
\label{SCA throughput OMA P1.5}{\rm{s.t.}}\;\;&\sum\nolimits_{n = 1}^N {\left( {\theta _k^l{{\left[ n \right]}^2} + 2\theta _k^l\left[ n \right]\left( {\theta _k^{}\left[ n \right] - \theta _k^l\left[ n \right]} \right)} \right)}  \ge \eta^{{\rm{O}}} ,\forall k,\\
\label{ep OMA P1.5}&
  {p_k}\left[ n \right] \le \chi _k^{lb}\left[ n \right],\forall n,k, \\
\label{w OMA 1.5}&\frac{{T{{\left[ n \right]}^4}}}{{\omega {{\left[ n \right]}^2}}} \!\le\! {\omega ^l}{\left[ n \right]^2} \!+\! 2{\omega ^l}\left[ n \right]\left( {\omega \left[ n \right] \!-\! {\omega ^l}\left[ n \right]} \right) \!+\! \frac{{s{{\left[ n \right]}^2}}}{{v_0^2}},\forall n,\\
\label{constraints OMA0}&\eqref{time duration OMA}-\eqref{transmit power OMA},\eqref{theta OMA P1.1},\eqref{UAV total energy OMA P1.1},\eqref{ep energy}.
\end{align}
\end{subequations}
It can be verified that Problem \eqref{P1.5} is a convex problem, which can be efficiently solved by standard convex optimization solvers such as CVX~\cite{cvx}. Similarly, the obtained objective value obtained from Problem \eqref{P1.5} serves a lower bound of that of Problem \eqref{P1.4} owing to the replacement of non-convex terms with their lower bounds.
\subsection{Overall Algorithm, Complexity, and Convergence}
Based on the two subproblems in the previous subsections, we propose an efficient algorithm to solve Problem \eqref{P1.1} by invoking AO method. Specifically, Problems \eqref{P1.3} and \eqref{P1.5} are alternately solved. The obtained solutions in each iteration are used as the input local points for the next iteration. The details of the proposed algorithm for OMA are summarized in \textbf{Algorithm 1}. The complexity of each subproblem with interior-point method are ${\mathcal{O}}\left( {{{\left( {3N + 2NK} \right)}^{3.5}}} \right)$ and ${\mathcal{O}}\left( {{{\left( {2N + 4NK} \right)}^{3.5}}} \right))$. Then, the total complexity for OMA is ${\mathcal{O}}\left( {N_{{\rm{ite}}}^{{\rm{O}}}\left( {{{\left( {5N + 6NK} \right)}^{3.5}}} \right)} \right)$, where $N_{{\rm{ite}}}^{{\rm{O}}}$ denotes the number of iterations needed for the convergence of \textbf{Algorithm 1}~\cite{convex}. It can be observed that the complexity of \textbf{Algorithm 1} is polynomial\footnote{Note that the proposed algorithm is still applicable to the case with a large number of GNs. This is because we consider an offline design and \textbf{Algorithm 1} is applied prior to the realistic UAV flight. Therefore, the potentially high complexity caused by large $K$ is acceptable given the available computing power.}.
\begin{remark}\label{remark:TDMA solution}
\emph{Although \textbf{Algorithm 1} is designed for solving the optimization problem with OMA-II, it is also applied for OMA-I with linear constraints: ${\tau _k}\left[ n \right] = {\tau _j}\left[ n \right],\forall k \ne j$.}
\end{remark}
Next, we demonstrate the convergence of \textbf{Algorithm 1}. The objective value of Problem \eqref{P1.1} in the $l$th iteration is defined as ${\eta ^{{\rm{O}}}}\left( {\left\{ {{{\mathbf{u}}^l}\left[ n \right]} \right\},\left\{ {{T^l}\left[ n \right]} \right\},\left\{ {\tau _k^l\left[ n \right]} \right\},\left\{ {p_k^l\left[ n \right]} \right\}} \right)$. First, for Problem \eqref{P1.3} with given transmit power in step 2 of \textbf{Algorithm 1}, we have
\begin{align}\label{P1.3 convergence}
\begin{gathered}
  \eta^{{\rm{O}}} \left( {\left\{ {{{\mathbf{u}}^l}\left[ n \right]} \right\},\left\{ {{T^l}\left[ n \right]} \right\},\left\{ {\tau _k^l\left[ n \right]} \right\},\left\{ {p_k^l\left[ n \right]} \right\}} \right) \hfill \\
  \mathop  = \limits^{\left( a \right)} \eta _p^{lb}\left( {\left\{ {{{\mathbf{u}}^l}\left[ n \right]} \right\},\left\{ {{T^l}\left[ n \right]} \right\},\left\{ {\tau _k^l\left[ n \right]} \right\},\left\{ {p_k^l\left[ n \right]} \right\}} \right) \hfill \\
  \mathop  \le \limits^{\left( b \right)} \eta _p^{lb}\left( {\left\{ {{{\mathbf{u}}^{l + 1}}\left[ n \right]} \right\},\left\{ {{T^{l + 0.5}}\left[ n \right]} \right\},\left\{ {\tau _k^{l + 0.5}\left[ n \right]} \right\},\left\{ {p_k^l\left[ n \right]} \right\}} \right) \hfill \\
  \mathop  \le \limits^{\left( c \right)} \eta^{{\rm{O}}} \left( {\left\{ {{{\mathbf{u}}^{l + 1}}\left[ n \right]} \right\},\left\{ {{T^{l + 0.5}}\left[ n \right]} \right\},\left\{ {\tau _k^{l + 0.5}\left[ n \right]} \right\},\left\{ {p_k^l\left[ n \right]} \right\}} \right), \hfill \\
\end{gathered}
\end{align}
where $\eta _p^{lb}$ represents the objective value of Problem \eqref{P1.3} with fixed transmit power. ${\left( a \right)}$ follows the fact that the first-order Taylor expansions are tight at the given local points in Problem \eqref{P1.3}; ${\left( b \right)}$ holds since Problem \eqref{P1.3} is solved optimally; ${\left( c \right)}$ holds due to the fact that the objective value of Problem \eqref{P1.3} serves the lower bound of that of \eqref{P1.2}. The inequality in \eqref{P1.3 convergence} suggests that the objective value of the original Problem \eqref{P1.2} is still non-decreasing after each iteration even if we only solve the approximate Problem \eqref{P1.3}.

Similarly, for Problem \eqref{P1.5} with given UAV location in step 3 of \textbf{Algorithm 1}, we have
\begin{align}\label{P1.5 convergence}
\begin{gathered}
  \eta^{{\rm{O}}} \left( {\left\{ {{{\mathbf{u}}^{l + 1}}\left[ n \right]} \right\},\left\{ {{T^{l + 0.5}}\left[ n \right]} \right\},\left\{ {t _k^{l + 0.5}\left[ n \right]} \right\},\left\{ {p_k^l\left[ n \right]} \right\}} \right) \hfill \\
  = \eta _{\mathbf{u}}^{lb}\left( {\left\{ {{{\mathbf{u}}^{l + 1}}\left[ n \right]} \right\},\left\{ {{T^{l + 0.5}}\left[ n \right]} \right\},\left\{ {t _k^{l + 0.5}\left[ n \right]} \right\},\left\{ {p_k^l\left[ n \right]} \right\}} \right) \hfill \\
  \le \eta _{\mathbf{u}}^{lb}\left( {\left\{ {{{\mathbf{u}}^{l + 1}}\left[ n \right]} \right\},\left\{ {{T^{l + 1}}\left[ n \right]} \right\},\left\{ {t _k^{l + 1}\left[ n \right]} \right\},\left\{ {p_k^{l+1}\left[ n \right]} \right\}} \right) \hfill \\
   \le \eta^{{\rm{O}}} \left( {\left\{ {{{\mathbf{u}}^{l + 1}}\left[ n \right]} \right\},\left\{ {{T^{l + 1}}\left[ n \right]} \right\},\left\{ {t _k^{l + 1}\left[ n \right]} \right\},\left\{ {p_k^{l + 1}\left[ n \right]} \right\}} \right), \hfill \\
\end{gathered}
\end{align}
where $\eta _{\mathbf{u}}^{lb}$ represents the objective value of Problem \eqref{P1.5} with fixed UAV location.\\
\begin{algorithm}[t!]\label{method0}
\caption{Proposed AO-based Algorithm for Solving Problem \eqref{P1.1}}
 \hspace*{0.02in}
\hspace*{0.02in} {Initialize feasible solutions $\left\{ {{{\mathbf{u}}^0}\left[ n \right],{T^0}\left[ n \right],\tau _k^0\left[ n \right],p_k^0\left[ n \right]} \right\}$ to \eqref{P1.1}, $l=0$.}\\
\vspace{-0.4cm}
\begin{algorithmic}[1]
\STATE {\bf repeat}
\STATE Solve Problem \eqref{P1.3} for given ${\left\{ {p_k^l\left[ n \right]} \right\}}$, and denote the optimal solutions as $\left\{ {{{\mathbf{u}}^{l + 1}}\left[ n \right],{T^{l + 0.5}}\left[ n \right],\tau _k^{l + 0.5}\left[ n \right]} \right\}$.
\STATE Solve Problem \eqref{P1.5} for given ${\left\{ {{{\mathbf{u}}^{l + 1}}\left[ n \right]} \right\}}$, and denote the optimal solutions as $\left\{ {p_k^{l + 1}\left[ n \right],{T^{l + 1}}\left[ n \right],\tau _k^{l + 1}\left[ n \right]} \right\}$.
\STATE $l=l+1$.
\STATE {\bf until} the fractional increase of the objective value is below a threshold $\xi   > 0$.
\end{algorithmic}
\end{algorithm}

As a result, based on \eqref{P1.3 convergence} and \eqref{P1.5 convergence}, we obtain that
\begin{align}\label{overall convergence P1.1}
\begin{gathered}
  \eta^{{\rm{O}}} \left( {\left\{ {{{\mathbf{u}}^l}\left[ n \right]} \right\},\left\{ {{T^l}\left[ n \right]} \right\},\left\{ {t _k^l\left[ n \right]} \right\},\left\{ {p_k^l\left[ n \right]} \right\}} \right) \hfill \\
   \le \eta^{{\rm{O}}} \left( {\left\{ {{{\mathbf{u}}^{l + 1}}\left[ n \right]} \right\},\left\{ {{T^{l + 1}}\left[ n \right]} \right\},\left\{ {t _k^{l + 1}\left[ n \right]} \right\},\left\{ {p_k^{l + 1}\left[ n \right]} \right\}} \right). \hfill \\
\end{gathered}
\end{align}
Equation \eqref{overall convergence P1.1} means the objective value of Problem \eqref{P1.1} is non-decreasing after each iteration. Since the max-min UAV data collection throughput is upper bounded by a finite value due to the limited energy at the UAV and GNs, the proposed algorithm is guaranteed to converge.
\section{Proposed Solution for NOMA}
To solve the formulated optimization problem for NOMA, we can use the same method to tackle the non-convex UAV energy constraint by introducing auxiliary variables $\left\{ {\omega \left[ n \right] \ge 0} \right\}$, as described in the previous section. To deal with other non-convexities of Problem \eqref{P2}, we first introduce auxiliary variables $\left\{ {{S_k}\left[ n \right]} \right\}$, $\left\{ {{I_k}\left[ n \right]} \right\}$, $\left\{ {{d_k}\left[ n \right]} \right\}$, and $\left\{ {{\theta _k}\left[ n \right]} \right\}$ such that
\begin{align}
\label{Skn}&{S_k}\left[ n \right] = \frac{{{{\left\| {{\mathbf{u}}\left[ n \right] - {{\mathbf{w}}_k}} \right\|}^2} + {H^2}}}{{{\gamma _0}{p_k}\left[ n \right]}},\forall k,n,\\
\label{Ikn}&{I_k}\left[ n \right] =   {\sum\nolimits_{m \in {{\mathcal{K}}},m \ne k} {{\gamma _0}{\alpha _{m,k}}\left[ n \right]{p_m}\left[ n \right]{d_m}{{\left[ n \right]}^{ - 1}}} }  + 1,\forall k,n,\\
\label{d NOMA}&{d_k}\left[ n \right] = {{{{\left\| {{\mathbf{u}}\left[ n \right] - {{\mathbf{w}}_k}} \right\|}^2} + {H^2}}},\forall k,n,\\
\label{theta NOMA}&{\theta _k}{\left[ n \right]^2} = \tau \left[ n \right]B{\log _2}\left( {1 + \frac{1}{{{S_k}\left[ n \right]{I_k}\left[ n \right]}}} \right),\forall k,n.
\end{align}
Define ${\eta ^{{\rm{N}}}} = \mathop {\min }\limits_{\forall k} Q_k^{{\rm{N}}}$, Problem \eqref{P2} can be equivalently rewritten as
\begin{subequations}\label{P2.1}
\begin{align}
&\mathop {\max }\limits_{\eta^{{\rm{N}}} ,\left\{ \begin{subarray}{l}
  {\mathbf{u}}\left[ n \right],T\left[ n \right],\tau \left[ n \right],{p_k}\left[ n \right],{\alpha _{k,m}}\left[ n \right] \\
  {S_k}\left[ n \right],{I_k}\left[ n \right],{\theta _k}\left[ n \right],\omega \left[ n \right],{d_k}\left[ n \right]
\end{subarray}  \right\}} \eta^{{\rm{N}}}  \\
\label{throughput NOMA P2.1}{\rm{s.t.}}\;\;&\sum\nolimits_{n = 1}^N {\theta _k}{\left[ n \right]^2}  \ge \eta^{{\rm{N}}} ,\forall k,\\
\label{theta NOMA P2.1}&\frac{{{\theta _k}{{\left[ n \right]}^2}}}{{\tau \left[ n \right]}} \le {\log _2}\left( {1 + \frac{1}{{{S_k}\left[ n \right]{I_k}\left[ n \right]}}} \right),\forall k,n,\\
\label{Skn 2.1}&{S_k}\left[ n \right] \ge \frac{{{{\left\| {{\mathbf{u}}\left[ n \right] - {{\mathbf{w}}_k}} \right\|}^2} + {H^2}}}{{{\gamma _0}{p_k}\left[ n \right]}},\forall k,n,\\
\label{Ikn 2.1}&\begin{gathered}
  {I_k}\left[ n \right] \ge  \hfill \\
  {\sum\nolimits_{m \in {{\mathcal{K}}},m \ne k} {{\gamma _0}{\alpha _{m,k}}\left[ n \right]{p_m}\left[ n \right]{d_m}{{\left[ n \right]}^{ - 1}}} }\!  +\! 1,\forall k,n, \hfill \\
\end{gathered} \\
\label{d NOMA 2.1}&{d_k}\left[ n \right] \le {{{{\left\| {{\mathbf{u}}\left[ n \right] - {{\mathbf{w}}_k}} \right\|}^2} + {H^2}}},\forall k,n,\\
\label{UAV total energy 2.1}&\begin{gathered}
  {P_0}\sum\nolimits_{n = 1}^N {\left( {T\left[ n \right] + \frac{{3s{{\left[ n \right]}^2}}}{{U_{tip}^2T\left[ n \right]}}} \right)}  + {P_i}\sum\nolimits_{n = 1}^N {\omega \left[ n \right]}  \hfill \\
  \;\;\;\;\;\;\;\;\;\;\;\;\;\;\;\;\; + \frac{1}{2}{d_0}\rho sA\sum\nolimits_{n = 1}^N {\frac{{s{{\left[ n \right]}^3}}}{{t{{\left[ n \right]}^2}}}}  \le {E_U}, \hfill \\
\end{gathered} \\
\label{w NOMA 2.1}&\frac{{T{{\left[ n \right]}^4}}}{{\omega {{\left[ n \right]}^2}}} \le \omega {\left[ n \right]^2} + \frac{{{{\left\| {{\mathbf{u}}\left[ {n + 1} \right] - {\mathbf{u}}\left[ n \right]} \right\|}^2}}}{{v_0^2}},\forall n,\\
\label{constraints NOMA}&\eqref{One UAV Initial Final Location Constraint NOMA},\eqref{One UAV Velocity Constraint NOMA},\eqref{sensor energy NOMA}-\eqref{index NOMA}.
\end{align}
\end{subequations}
The equivalence between \eqref{P2} and \eqref{P2.1} can be shown similarly as \textbf{Proposition 1}. It is observed that Problem \eqref{P2.1} has a similar structure with Problem \eqref{P1.1} except integer constraints. Therefore, we still decompose \eqref{P2.1} into several subproblems, which are ease to handle.
\subsection{Optimization with Fixed Transmit Power and Decoding Order}
For any given feasible GN transmit power, $\left\{ {{p_k}\left[ n \right]} \right\}$, and decoding orders, $\left\{ {{a_{k,m}}\left[ n \right]} \right\}$, the optimization problem can be written as
\begin{subequations}\label{P2.2}
\begin{align}
&\mathop {\max }\limits_{\eta^{{\rm{N}}} ,\left\{ \begin{subarray}{l}
  {\mathbf{u}}\left[ n \right],T\left[ n \right],\tau \left[ n \right],{S_k}\left[ n \right] \\
  {I_k}\left[ n \right],{\theta _k}\left[ n \right],\omega \left[ n \right],{d_k}\left[ n \right]
\end{subarray}  \right\}} \eta^{{\rm{N}}}  \\
\label{constraints P2.2}{\rm{s.t.}}\;\;&\eqref{One UAV Initial Final Location Constraint NOMA},\eqref{One UAV Velocity Constraint NOMA},\eqref{sensor energy NOMA},\eqref{time duration NOMA},\eqref{throughput NOMA P2.1}-\eqref{w NOMA 2.1}.
\end{align}
\end{subequations}
Problem \eqref{P2.2} is still non-convex owing to the non-convex constraints \eqref{throughput NOMA P2.1}, \eqref{theta NOMA P2.1}, \eqref{d NOMA 2.1}, and \eqref{w NOMA 2.1}. Specifically, \eqref{throughput NOMA P2.1} and \eqref{w NOMA 2.1} can be handled as introduced in the previous section. Before handling the non-convex constraint \eqref{theta NOMA P2.1}, we first have the following lemma.
\begin{lemma}\label{lemma:xy}
\emph{For $x > 0$ and $y > 0$, $f\left( {x,y} \right) = {\log _2}\left( {1 + \frac{1}{{xy}}} \right)$ is a joint convex function with respect to $x$ and $y$}
\begin{proof}
\emph{Lemma \ref{lemma:xy} can be proved by showing the Hessian matrix of function $f\left( {x,y} \right)$ is positive semidefinite when $x > 0$ and $y > 0$. As a result, $f\left( {x,y} \right)$ is a convex function.}
\end{proof}
\end{lemma}
Based on \textbf{Lemma \ref{lemma:xy}}, the RHS of \eqref{theta NOMA P2.1} is jointly convex with respect to ${{S_k}\left[ n \right]}$ and ${{I_k}\left[ n \right]}$. Thus, by applying the first-order Taylor explanation, the lower bound at given local points ${\left\{ {S_k^l\left[ n \right],I_k^l\left[ n \right]} \right\}}$ can be expressed as
\begin{align}\label{Rkn NOMA lower bound}
\begin{gathered}
   {\log _2}\left( {1 \!+\! \frac{1}{{{S_k}\left[ n \right]{I_k}\left[ n \right]}}} \right) \!\ge\! R_k^{lb}\left[ n \right]\! =\! \log_2 \left( {1 \!+\! \frac{1}{{S_k^l\left[ n \right]I_k^l\left[ n \right]}}} \right) \hfill \\
   - \frac{{\left( {{{\log }_2}e} \right)\left( {S_k^{}\left[ n \right] - S_k^l\left[ n \right]} \right)}}{{S_k^l\left[ n \right] + S_k^l{{\left[ n \right]}^2}I_k^l\left[ n \right]}} - \frac{{\left( {{{\log }_2}e} \right)\left( {I_k^{}\left[ n \right] - I_k^l\left[ n \right]} \right)}}{{I_k^l\left[ n \right] + I_k^l{{\left[ n \right]}^2}S_k^l\left[ n \right]}}. \hfill \\
\end{gathered}
\end{align}

Furthermore, for the non-convex constraints \eqref{d NOMA 2.1}, the RHS is a convex function with respect to ${{\mathbf{u}}\left[ n \right]}$. The corresponding lower bound at given local points ${{{\mathbf{u}}^l}\left[ n \right]}$ is expressed as
\begin{align}\label{d lower bound}
\begin{gathered}
  {\left\| {{\mathbf{u}}\left[ n \right] - {{\mathbf{w}}_k}} \right\|^2} \ge {\left\| {{{\mathbf{u}}^l}\left[ n \right] - {{\mathbf{w}}_k}} \right\|^2} \hfill \\
  \;\;\;\;\;\;\;\;\;\;\;\;\;\;\;\;\;\;\;\;\;\;\;\;\; + 2{\left( {{{\mathbf{u}}^l}\left[ n \right] - {{\mathbf{w}}_k}} \right)^T}\left( {{{\mathbf{u}}}\left[ n \right] - {{\mathbf{u}}^l}\left[ n \right]} \right). \hfill \\
\end{gathered}
\end{align}
Therefore, Problem \eqref{P2.2} is approximated as the following optimization problem:
\begin{subequations}\label{P2.3}
\begin{align}
&\mathop {\max }\limits_{\eta^{{\rm{N}}} ,\left\{ \begin{subarray}{l}
  {\mathbf{u}}\left[ n \right],T\left[ n \right],\tau \left[ n \right],{S_k}\left[ n \right] \\
  {I_k}\left[ n \right],{\theta _k}\left[ n \right],\omega \left[ n \right],{d_k}\left[ n \right]
\end{subarray}  \right\}} \eta^{{\rm{N}}}  \\
\label{SCA throughput NOMA 2.3}{\rm{s.t.}}\;\;&\sum\nolimits_{n = 1}^N {\left( {\theta _k^l{{\left[ n \right]}^2} + 2\theta _k^l\left[ n \right]\left( {\theta _k^{}\left[ n \right] - \theta _k^l\left[ n \right]} \right)} \right)}  \ge \eta^{{\rm{N}}} ,\forall k,\\
\label{SCA theta NOMA 2.3}&\frac{{{\theta _k}{{\left[ n \right]}^2}}}{{{\tau _k}\left[ n \right]}} \le R_k^{lb}\left[ n \right],\forall k,n,\\
\label{d P2.3}&\begin{gathered}
  {d_k}\left[ n \right] \le {H^2} + {\left\| {{{\mathbf{u}}^l}\left[ n \right] - {{\mathbf{w}}_k}} \right\|^2} \hfill \\
  \;\;\;\;\;\;\;\;\;\;\;\;\;\; + 2{\left( {{{\mathbf{u}}^l}\left[ n \right] - {{\mathbf{w}}_k}} \right)^T}\left( {{{\mathbf{u}}}\left[ n \right] - {{\mathbf{u}}^l}\left[ n \right]} \right), \hfill \\
\end{gathered}\\
\label{w NOMA 2.3}&\frac{{T{{\left[ n \right]}^4}}}{{\omega {{\left[ n \right]}^2}}} \le {\beta ^{lb}}\left[ n \right],\forall n,\\
\label{constraints NOMA 2.3}&\eqref{One UAV Initial Final Location Constraint NOMA},\eqref{One UAV Velocity Constraint NOMA},\eqref{sensor energy NOMA},\eqref{time duration NOMA},\eqref{Skn 2.1},\eqref{Ikn 2.1},\eqref{UAV total energy 2.1}.
\end{align}
\end{subequations}
Problem \eqref{P2.3} is a convex problem that can be efficiently solved by standard convex optimization solvers such as CVX~\cite{cvx}. The optimal objective value obtained from Problem \eqref{P2.3} provides a lower bound to that of Problem \eqref{P2.2}.
\subsection{Optimization with Fixed UAV Locations and Decoding Order}
For any given feasible UAV location ${\left\{ {{\mathbf{u}}\left[ n \right]} \right\}}$ and decoding orders $\left\{ {{a_{k,m}}\left[ n \right]} \right\}$, Problem \eqref{P2.1} with auxiliary variables $\left\{ {\varepsilon \left[ n \right]} \right\}$ can be written as the following optimization problem
\begin{subequations}\label{P2.4}
\begin{align}
&\mathop {\max }\limits_{\eta^{{\rm{N}}} ,\left\{ \begin{subarray}{l}
  T\left[ n \right],\tau \left[ n \right],{p_k}\left[ n \right],{S_k}\left[ n \right] \\
  {I_k}\left[ n \right],{\theta _k}\left[ n \right],\omega \left[ n \right],{\varepsilon \left[ n \right]},{d_k}\left[ n \right]
\end{subarray}  \right\}} \eta^{{\rm{N}}}  \\
\label{ep energy NOMA}{\rm{s.t.}}\;\;&\sum\nolimits_{n = 1}^N {{\varepsilon _k}{{\left[ n \right]}^2}}  \le {E_k},\forall k,\\
\label{ep constraint NOMA}&{p_k}\left[ n \right] \le \frac{{{\varepsilon _k}{{\left[ n \right]}^2}}}{{{\tau }\left[ n \right]}},\forall n,k,\\
\label{P2.4 constraints}&\eqref{time duration NOMA},\eqref{transmit power NOMA},\eqref{throughput NOMA P2.1}-\eqref{w NOMA 2.1}.
\end{align}
\end{subequations}
By replacing those non-convex terms involved in Problem \eqref{P2.4} with their lower bounds, \eqref{P2.4} is approximated as the following problem
\begin{subequations}\label{P2.5}
\begin{align}
&\mathop {\max }\limits_{\eta^{{\rm{N}}} ,\left\{ \begin{subarray}{l}
  T\left[ n \right],\tau \left[ n \right],{p_k}\left[ n \right],{S_k}\left[ n \right] \\
  {I_k}\left[ n \right],{\theta _k}\left[ n \right],\omega \left[ n \right],{\varepsilon \left[ n \right]},{d_k}\left[ n \right]
\end{subarray}  \right\}} \eta^{{\rm{N}}}  \\
\label{SCA throughput NOMA 2.5}{\rm{s.t.}}\;\;&\sum\nolimits_{n = 1}^N {\left( {\theta _k^l{{\left[ n \right]}^2} + 2\theta _k^l\left[ n \right]\left( {\theta _k^{}\left[ n \right] - \theta _k^l\left[ n \right]} \right)} \right)}  \ge \eta^{{\rm{N}}} ,\forall k,\\
\label{SCA theta NOMA 2.5}&\frac{{{\theta _k}{{\left[ n \right]}^2}}}{{{\tau }\left[ n \right]}} \le R_k^{lb}\left[ n \right],\forall k,n,\\
\label{ep NOMA 2.5}&
  {p_k}\left[ n \right] \le \chi _k^{lb}\left[ n \right],\forall n,k, \\
\label{w NOMA 2.5}&\frac{{T{{\left[ n \right]}^4}}}{{\omega {{\left[ n \right]}^2}}} \le {\omega ^l}{\left[ n \right]^2} \!+\! 2{\omega ^l}\left[ n \right]\left( {\omega \left[ n \right] \!-\! {\omega ^l}\left[ n \right]} \right) \!+\! \frac{{s{{\left[ n \right]}^2}}}{{v_0^2}},\forall n,\\
\label{P2.5 constraints}&\eqref{time duration NOMA},\eqref{transmit power NOMA},\eqref{Skn 2.1}-\eqref{UAV total energy 2.1}.
\end{align}
\end{subequations}
Here, the expression of $\chi _k^{lb}\left[ n \right]$ is obtained by dropping the index $k$ of $\tau _k^l\left[ n \right]$ and $\tau _k\left[ n \right]$ in \eqref{ep lower bound}. Then, Problem \eqref{P2.5} is a convex problem that can be efficiently solved by standard convex optimization solvers such as CVX~\cite{cvx}, and the optimal objective value obtained from Problem \eqref{P2.5} serves a lower bound of that of Problem \eqref{P2.4}.
\subsection{Decoding Order Design with Other Variables Fixed}
For any given feasible UAV trajectory, $\left\{ {{\mathbf{u}}\left[ n \right],T\left[ n \right]} \right\}$, the time allocation, $\left\{ {\tau \left[ n \right]} \right\}$, and the GN transmit power, $\left\{ {{p_k}\left[ n \right]} \right\}$, the optimization problem \eqref{P2.1} is reduced to
\begin{subequations}\label{P2.6}
\begin{align}
&\mathop {\max }\limits_{\eta^{{\rm{N}}} ,\left\{ {{a_{k,m}}\left[ n \right],{I_k}\left[ n \right]} \right\}} \eta^{{\rm{N}}} \\
\label{sum throughput NOMA P2.6}{\rm{s.t.}}\;\;&\sum\nolimits_{n = 1}^N {\tau \left[ n \right]{{\log }_2}\left( {1 + \frac{1}{{{S_k}\left[ n \right]{I_k}\left[ n \right]}}} \right) \ge \eta^{{\rm{N}}} } ,\forall k,\\
\label{P2.6 constraints}&\eqref{decoding order NOMA},\eqref{index NOMA},\eqref{Ikn 2.1}.
\end{align}
\end{subequations}
Involving integer constraints \eqref{index NOMA} and non-convex constraint \eqref{sum throughput NOMA P2.6}, Problem \eqref{P2.6} is a mixed integer non-convex optimization problem. The integer constraint \eqref{index NOMA} can be equivalently transformed into the following two constraints:
\begin{align}
\label{integer 1}&\sum\nolimits_{k = 1}^K {\sum\nolimits_{m \ne k}^K {\left( {{\alpha _{k,m}}{{\left[ n \right]}^2} - {\alpha _{k,m}}\left[ n \right]} \right)} }  \ge 0,\\
\label{integer 2}&0 \le {\alpha _{k,m}}\left[ n \right] \le 1,\forall k \ne m \in \mathcal{K}.
\end{align}
Consequently, Problem \eqref{P2.6} can be reformulated with continuous variables $\left\{ {{\alpha _{k,m}}\left[ n \right]} \right\}$ as
\begin{subequations}\label{P2.7}
\begin{align}
&\mathop {\max }\limits_{\eta^{{\rm{N}}} ,\left\{ {{a_{k,m}}\left[ n \right],{I_k}\left[ n \right]} \right\}} \eta^{{\rm{N}}} \\
\label{P2.7 constraints}{\rm{s.t.}}\;\;&\eqref{decoding order NOMA},\eqref{Ikn 2.1},\eqref{sum throughput NOMA P2.6},\eqref{integer 1},\eqref{integer 2}.
\end{align}
\end{subequations}
Though removing integer constraints, Problem \eqref{P2.7} is still a non-convex problem with non-convex constraints \eqref{sum throughput NOMA P2.6} and \eqref{integer 1}. Before handling Problem \eqref{P2.7}, we first have the following theorem.
\begin{theorem}\label{penlty}
\emph{For a sufficiently large constant value $\lambda  \gg 1$, Problem \eqref{P2.7} is equivalent to the following problem
\begin{subequations}\label{P2.8}
\begin{align}
\mathop {\max }\limits_{\eta^{{\rm{N}}} ,\left\{ {{a_{k,m}}\left[ n \right],{I_k}\left[ n \right]} \right\}}& \eta^{{\rm{N}}}  \!\!+\! \lambda\! \sum\nolimits_{k = 1}^K \!{\sum\nolimits_{m \ne k}^K {\!\left(\! {{\alpha _{k,m}}{{\left[ n \right]}^2} \!-\! {\alpha _{k,m}}\left[ n \right]} \!\right)\!} }  \\
\label{P2.8 constraints}{\rm{s.t.}}\;\;&\eqref{decoding order NOMA},\eqref{Ikn 2.1},\eqref{sum throughput NOMA P2.6},\eqref{integer 2}.
\end{align}
\end{subequations}
where $\lambda $ represents a penalty factor to penalize the objective function for any ${\alpha _{k,m}}\left[ n \right]$ that is not equal to 0 or 1.
}
\begin{proof}
See Appendix~A.
\end{proof}
\end{theorem}
To handle Problem \eqref{P2.8}, we only need to deal with the non-convex objective function \eqref{P2.8} and the non-convex constraint \eqref{sum throughput NOMA P2.6}. The second term of \eqref{P2.8} is a convex function with respect to ${\alpha _{k,m}}\left[ n \right]$. By utilizing the first-order Taylor expansion at a given local point $\alpha _{k,m}^l\left[ n \right]$, the lower bound is expressed as
\begin{align}\label{a lower bound}
  {\alpha _{k,m}}{\!\left[ n \right]^2} \!\!\ge\!\!  {\xi _{k,m}}\!\left[ n \right] \!\!=\!\! \alpha _{k,m}^l{\!\left[ n \right]^2} \!\!+\!\! 2\alpha _{k,m}^l\!\left[ n \right]\left( {\alpha _{k,m}^{}\!\left[ n \right] \!\!-\!\! \alpha _{k,m}^l\!\left[ n \right]} \right).
\end{align}
For the non-convex constraint \eqref{sum throughput NOMA P2.6}, the LHS is a convex function with respect to ${{I_k}\left[ n \right]}$. Similarly, the lower bound at given local points ${\left\{ {I_k^l\left[ n \right]} \right\}}$ is expressed as
\begin{align}\label{Rkn low decoding order}
\begin{gathered}
  {\log _2}\left( {1 + \frac{1}{{{S_k}\left[ n \right]{I_k}\left[ n \right]}}} \right) \ge \mu _k^{lb}\left[ n \right] \hfill \\
   = {\log _2}\left( {1 + \frac{1}{{{S_k}\left[ n \right]I_k^l\left[ n \right]}}} \right) - \frac{{\left( {{{\log }_2}e} \right)\left( {{I_k}\left[ n \right] - I_k^l\left[ n \right]} \right)}}{{I_k^l\left[ n \right] + {S_k}\left[ n \right]I_k^l{{\left[ n \right]}^2}}}. \hfill \\
\end{gathered}
\end{align}
Based on \eqref{a lower bound} and \eqref{Rkn low decoding order}, Problem \eqref{P2.8} can be approximated as the following problem:
\begin{subequations}\label{P2.9}
\begin{align}
\mathop {\max }\limits_{\eta^{{\rm{N}}} ,\left\{ {{a_{k,m}}\left[ n \right],{I_k}\left[ n \right]} \right\}}& \eta^{{\rm{N}}}  \!+\! \lambda\! \sum\nolimits_{k = 1}^K \!{\sum\nolimits_{m \ne k}^K {\left( {{\xi _{k,m}}\left[ n \right] \!-\! {\alpha _{k,m}}\left[ n \right]} \right)} } \\
\label{sum throughput NOMA P2.9}{\rm{s.t.}}\;\;&
  \sum\nolimits_{n = 1}^N {{\tau}\left[ n \right]}\mu _k^{lb}\left[ n \right] \ge \eta^{{\rm{N}}} ,\forall k,\\
\label{P2.9 constraints}&\eqref{decoding order NOMA},\eqref{Ikn 2.1},\eqref{sum throughput NOMA P2.6},\eqref{integer 2}.
\end{align}
\end{subequations}
Problem \eqref{P2.9} is a convex problem that can be solved efficiently by standard convex program solvers such as CVX~\cite{cvx}. Specifically, we develop an iterative algorithm to optimize the decoding orders for given UAV trajectory, time allocation, and transmit power as summarized in \textbf{Algorithm 2}. Since the application of lower bounds approximation, the obtained result serves a lower bound of that of Problem \eqref{P2.8}.
\begin{algorithm}[!t]\label{method2}
\caption{Proposed Penalty-based Algorithm for Solving Problem \eqref{P2.9}}
 \hspace*{0.02in}
\hspace*{0.02in} {Initialize the penalty factor $\lambda $ and feasible solutions $\left\{ {\alpha _{k,m}^l\left[ n \right]} \right\}$ to \eqref{P2.8} with given $\left\{ {{{\mathbf{u}}}\left[ n \right],{T}\left[ n \right],\tau \left[ n \right],p_k\left[ n \right]} \right\}$, $l=0$.}\\
\vspace{-0.4cm}
\begin{algorithmic}[1]
\STATE {\bf repeat}
\STATE Solve Problem \eqref{P2.9} with $\left\{ {\alpha _{k,m}^l\left[ n \right]} \right\}$, and denote optimal solutions as $\left\{ {\alpha _{k,m}^{l + 1}\left[ n \right]} \right\}$.
\STATE $l=l+1$.
\STATE {\bf until} the fractional increase of the objective value is below a threshold $\xi   > 0$.
\end{algorithmic}
\end{algorithm}
\subsection{Overall Algorithm, Complexity, and Convergence}
Based on subproblems in the previous three subsections, we propose an efficient iterative algorithm to solve Problem \eqref{P2} by invoking AO method. The details of the designed algorithm for NOMA are summarized in \textbf{Algorithm 3}. The complexity of each subproblem with interior-point method are ${\mathcal{O}}\left( {{{\left( {4N + 4NK} \right)}^{3.5}}} \right)$, ${\mathcal{O}}\left( {{{\left( {4N + 5NK} \right)}^{3.5}}} \right)$, and ${\mathcal{O}}\left( {N_{{\rm{ite}}}^2{{\left( {N{K^2} + NK} \right)}^{3.5}}} \right)$, respectively, where $N_{{\rm{ite}}}^2$ is the number of iterations required for the convergence of \textbf{Algorithm 2}. Therefore, the total complexity for NOMA is ${\mathcal{O}}\left( {N_{{\rm{ite}}}^{{\rm{N}}}\left( {{{\left( {8N + 9NK} \right)}^{3.5}} + N_{{\rm{ite}}}^2{{\left( {N{K^2} + NK} \right)}^{3.5}}} \right)} \right)$, where $N_{{\rm{ite}}}^{{\rm{N}}}$ is the iteration number of \textbf{Algorithm 3} for NOMA~\cite{convex}. It can be seen that complexity of NOMA is larger than that of OMA. The convergency of \textbf{Algorithm 3} can be shown similarly as that of \textbf{Algorithm 1}. The details are omitted for brevity.
\begin{algorithm}[t!]\label{method1}
\caption{Proposed AO-based Algorithm for Solving Problem \eqref{P2}}
 \hspace*{0.02in}
\hspace*{0.02in} {Initialize feasible solutions to \eqref{P1}\\ $\left\{ {{{\mathbf{u}}^0}\left[ n \right],{T^0}\left[ n \right],\tau^0 \left[ n \right],p_k^0\left[ n \right],{\alpha _{k,m}^0}\left[ n \right]} \right\}$. $l=0$.}\\
\vspace{-0.4cm}
\begin{algorithmic}[1]
\STATE {\bf repeat}
\STATE Solve Problem \eqref{P2.3} for given $\left\{ {p_k^l\left[ n \right],a_{k,m}^l\left[ n \right]} \right\}$, and denote the optimal solutions by $\left\{ {{{\mathbf{u}}^{l + 1}}\left[ n \right],{T^{l+0.5}}\left[ n \right],\tau ^{l+0.5}\left[ n \right]} \right\}$.
\STATE Solve Problem \eqref{P2.5} for given $\left\{ {{{\mathbf{u}}^{l + 1}}\left[ n \right],a_{k,m}^l\left[ n \right]} \right\}$, and denote the optimal solutions by $\left\{ {p_k^{l + 1}\left[ n \right],{T^{l + 1}}\left[ n \right],\tau ^{l + 1}\left[ n \right]} \right\}$.
\STATE Solve Problem \eqref{P2.9} for given $\left\{ {{{\mathbf{u}}^{l + 1}}\left[ n \right],{T^{l + 1}}\left[ n \right],\tau ^{l + 1}\left[ n \right],p_k^{l + 1}\left[ n \right]} \right\}$ via \textbf{Algorithm 2}. Denote the optimal solutions by $\left\{ {\alpha _{k,m}^{l + 1}\left[ n \right]} \right\}$.
\STATE $l=l+1$.
\STATE {\bf until} the fractional increase of the objective value is below a threshold $\xi    > 0$.
\end{algorithmic}
\end{algorithm}
\section{Numerical Examples}
\begin{figure}[t!]
\centering
\subfigure[$E_U=8$ KJ.]{\label{trajectory 8KJ}
\includegraphics[width= 2.5in]{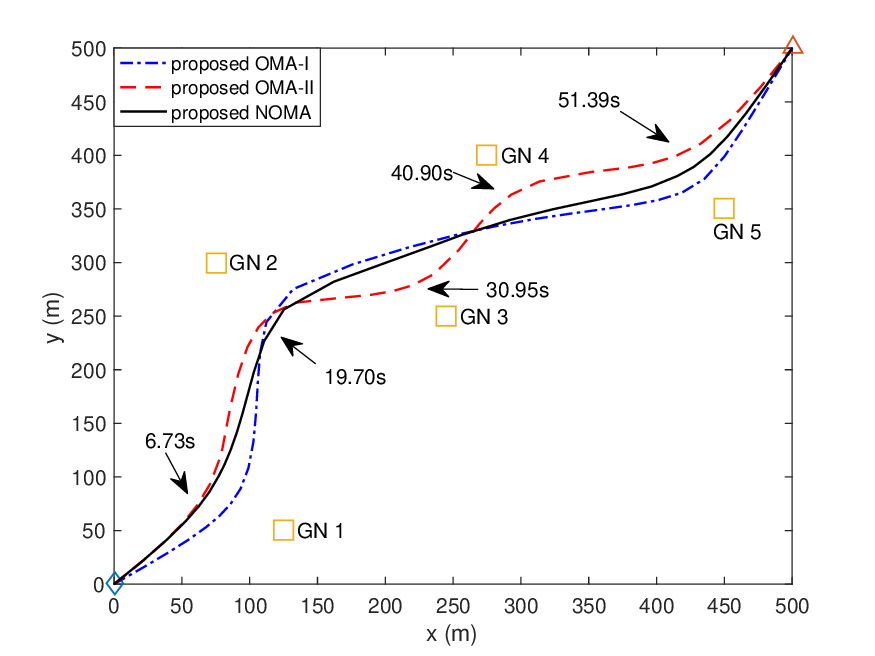}}
\subfigure[$E_U=30$ KJ.]{\label{trajectory 50KJ}
\includegraphics[width= 2.5in]{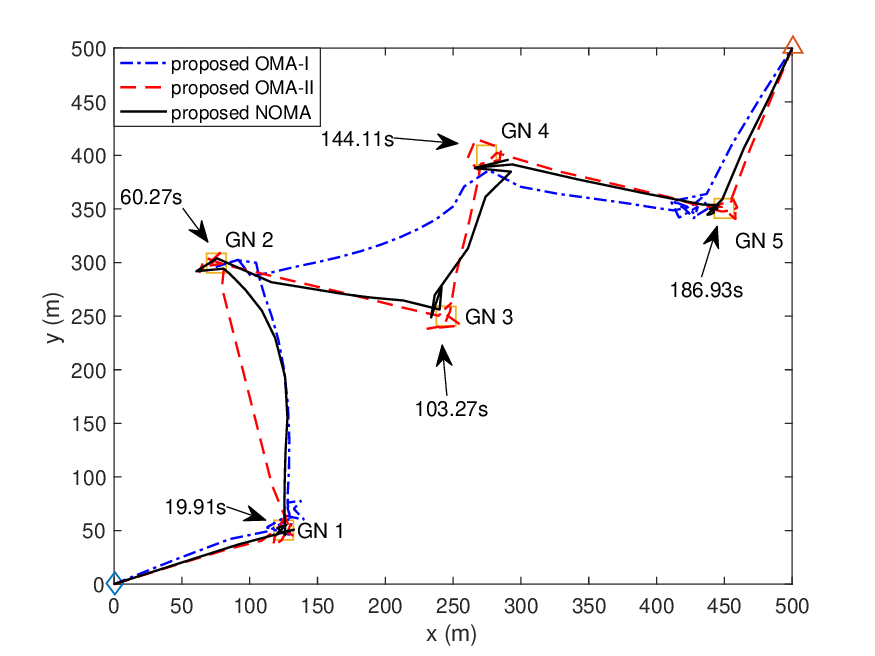}}
\setlength{\abovecaptionskip}{-0cm}
\caption{The optimized UAV trajectories for differen MA schemes and $E_s=10$ J.}\label{UAV trajectories}
\end{figure}

In this section, numerical examples are provided to evaluate the performances of the proposed algorithms. In the simulations, we consider a UAV data collection system with $K=5$ GNs, which are randomly and uniformly distributed in a square area of $500 \times 500\;{{{\rm{m}}}^2}$. As the maximum allowed UAV height in federal aviation authority (FAA) regulations is 122m, the height of the UAV is fixed at $H=100$ m to make the A2G channel can be well approximated by the LoS channel model. The UAV is assumed to fly from the initial location ${\left( {0,0,100} \right)^T}$ m to the final location ${\left( {500,500,100} \right)^T}$ m. The maximum UAV speed is ${V_{\max }} = 30$ m/s. For the rotary-wing UAV propulsion power consumption model in \eqref{UAV Power Model}, we set the parameters as follows\footnote{As reported in~[Table I, 16], the parameters for the rotary-wing UAV propulsion power consumption model are selected in terms of the physical model of the UAV (e.g., UAV weight, rotor size, blade angular, etc) and the objective environment (i.e., air density).}~\cite{Zeng2019Energy}: ${P_0}=79.86$ W, $P_i=88.63$ W, $U_{tip}=120$ m/s, $v_0=4.03$ m/s, $d_0=0.6$, $\rho = 1.225 $ kg/${{\rm{m}}}^3$, $s=0.05$, $A=0.503$ ${{\rm{m}}}^2$. The received SNR at a reference distance of $1$ m is ${\gamma _0} =  50$ dB. The maximum transmit power of GNs is set to $P_{\max}=0.1$ W. As the max-min problem is considered to guarantees the fairness among GNs, without lose of generality, we assume that all GNs have identical storage energy (i.e., ${E_k} = {E_s},\forall k$) to ensure them in a fair initial state. The algorithm threshold $\xi $ is set to ${10^{ - 2}}$. The following results are obtained based on one random GN deployment as illustrated in Fig. \ref{UAV trajectories} via changing different parameters (e.g., UAV on-board energy and GN storage energy).

\begin{figure}[t!]
    \begin{center}
        \includegraphics[width=2.5in]{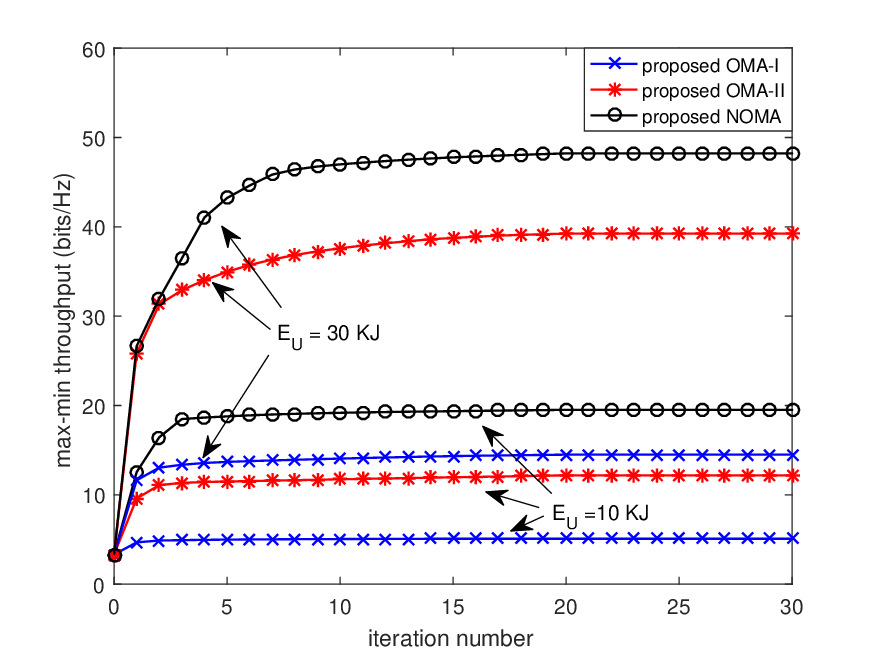}
        \caption{Convergence of the proposed algorithms.}
        \label{Convergence}
    \end{center}
\end{figure}

In Fig. \ref{Convergence}, we first study the convergence of \textbf{Algorithm 1} and \textbf{Algorithm 3} for OMA and NOMA cases with $E_s=10$ Joule (J). The initial UAV trajectory, $\left\{ {{{\mathbf{u}}^0}\left[ n \right],{T^0}\left[ n \right]} \right\}$, is set to the straight flight from the initial location to the final location with the maximum-range (MR) speed in~\cite{Zeng2019Energy}. For OMA, the initial time allocation, $\left\{ {\tau _k^0\left[ n \right]} \right\}$, and initial transmit power, $\left\{ {p_k^0\left[ n \right]} \right\}$, are obtained by letting $\tau _k^0\left[ n \right] = \frac{{{T^0}\left[ n \right]}}{K}$ and $p_k^0\left[ n \right] = \min \left( {{P_{\max }},{{{E_s}} \mathord{\left/
 {\vphantom {{{E_s}} {\sum\nolimits_{n = 1}^N {\tau _k^0\left[ n \right]} }}} \right.
 \kern-\nulldelimiterspace} {\sum\nolimits_{n = 1}^N {\tau _k^0\left[ n \right]} }}} \right),\forall k,n$. For NOMA, the initial time allocation, $\left\{ {\tau ^0\left[ n \right]} \right\}$, and initial transmit power, $\left\{ {p_k^0\left[ n \right]} \right\}$, are obtained by letting ${\tau ^0}\left[ n \right] = {T^0}\left[ n \right]$ and $p_k^0\left[ n \right] = \min \left( {{P_{\max }},{{{E_s}} \mathord{\left/
 {\vphantom {{{E_s}} {\sum\nolimits_{n = 1}^N {\tau _k^0\left[ n \right]} }}} \right.
 \kern-\nulldelimiterspace} {\sum\nolimits_{n = 1}^N {\tau ^0\left[ n \right]} }}} \right),\forall k,n$. We consider two cases with $E_U=10$ KJ and $E_U=30$ KJ. From the figure, it is observed that the max-min achievable throughput of three schemes increase as the number of iterations increases. When $E_U=10$ KJ, the proposed algorithm for three schemes converges with around 10 iterations. When $E_U=30$ KJ, the proposed algorithm converges with around 25 iterations. The reasons behind this can be explained as follows. Since $E_U=30$ KJ enables more degrees-of-freedom for UAV trajectory design than $E_U=10$ KJ, the UAV for large $E_U$ can achieve more complex trajectory to collect more information bits than that for small $E_U$ (which are shown in the following optimize UAV trajectories of Fig. \ref{UAV trajectories}). However, recall the fact that the initial UAV trajectory is same for different $E_U$ and the employed path discretization method makes the displacement of the UAV at each iteration is relatively small. Therefore, the proposed algorithm for $E_U=30$ KJ needs more iterations to converge than that for $E_U=10$ KJ.

\begin{figure}[t!]
\centering
\subfigure[$E_U=8$ KJ.]{\label{speed 8KJ}
\includegraphics[width= 2.5in]{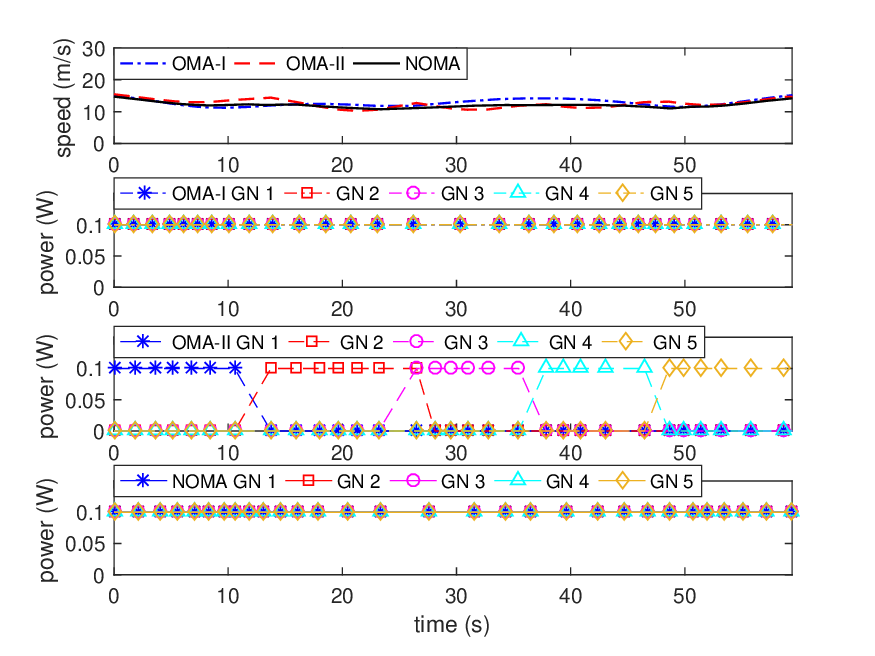}}
\subfigure[$E_U=30$ KJ.]{\label{speed 50KJ}
\includegraphics[width= 2.5in]{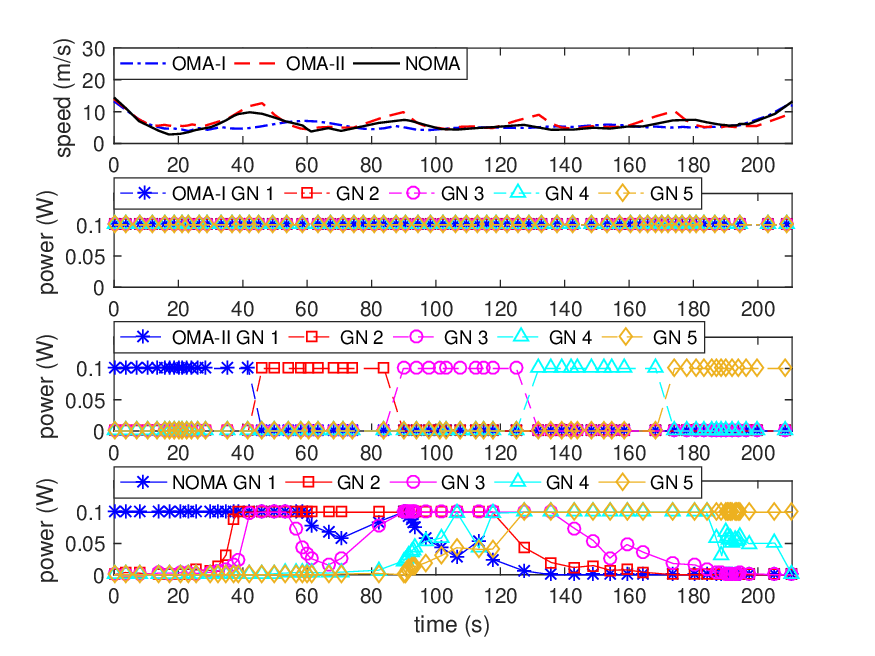}}
\setlength{\abovecaptionskip}{-0cm}
\caption{The optimized speed of the UAV and transmit power of GNs for differen MA schemes and $E_s=10$ J.}\label{speed and power}
\end{figure}

In Fig. \ref{UAV trajectories}, we provide the optimized UAV trajectory for different MA schemes with $E_s=10$ J and different $E_U$. In order to illustrate the variety of the instant UAV speed and GNs' transmit power shown in Fig. \ref{speed and power}, Fig. \ref{UAV trajectories} also presents the time instant when the UAV is closest to each GN in the OMA-II scheme. It is first observed from Fig. \ref{UAV trajectories} that the UAV tries to successively fly as close as possible to each GN in both three schemes even with different $E_U$. This is expected since the considered max-min throughput objective function makes the UAV need to collect data from each GN in a fair manner. When the UAV on-board energy is small, e.g., $E_U=8$ KJ, the obtained UAV paths and speeds for three schemes are similar, as shown in Fig. \ref{trajectory 8KJ} and Fig. \ref{speed 8KJ}. Regarding GNs' transmit power, GNs in all schemes tend to transmit at $P_{\max}$ as long as being waken up by the UAV. This is because, in this case, $E_s$ is large enough compared with the total available communication time. Note that GNs transmit in a successive manner for OMA-II. This is because the optimization of time resources allocation in OMA-II allows the UAV to allocate all the communication time resources to its nearest GN along the trajectory, thus collecting more information bits. This phenomena can be also verified by the time instant in Fig. \ref{UAV trajectories} and the corresponding GN awake state in Fig. \ref{speed and power}. Though all GNs transmit at $P_{\max}$ through the whole UAV flight time in NOMA and OMA-I, the reasons are different. For NOMA, all GNs are multiplexed in power levels within the same time/frequency resources. As a result, the UAV can not only wake up its nearest GNs but also other GNs to collect more data. However, for OMA-I, as the time resources are always equally allocated to GNs, GNs need to stay awake throughout the whole time to upload more data to the UAV.

In Fig. \ref{trajectory 50KJ} and Fig. \ref{speed 50KJ} for $E_U=30$ KJ, the UAV in general successively flies to the top of each GN in both schemes. It is observed that the UAV keeps flying around at the top of GNs other than remaining static. The reason is that the UAV will consume much higher energy to hover in the air than flying around with a certain speed. Therefore, the saved energy by flying around instead of hovering can prolong the communication time and increase the achieved throughput. Moreover, GNs in NOMA do not always keep awake when $E_U=30$ KJ. This is expected since the UAV flight time for $E_U=30$ KJ is rather large, the storage energy $E_s$ is not enough to allow GNs to keep transmitting. As a result, the UAV tends to wake up some of GNs to get full use of the limited energy stored at each GN. For OMA-I with $E_U=30$ KJ, GNs still keep awake during the UAV flight time due to the equal time allocation. It also causes the UAV for NOMA and OMA-I in Fig. \ref{trajectory 50KJ} flies a curve between two GNs instead of a straight line for OMA-II since the UAV tends to maximize the rate of all awake GNs while flying.
\begin{figure}[t!]
    \begin{center}
        \includegraphics[width=2.5in]{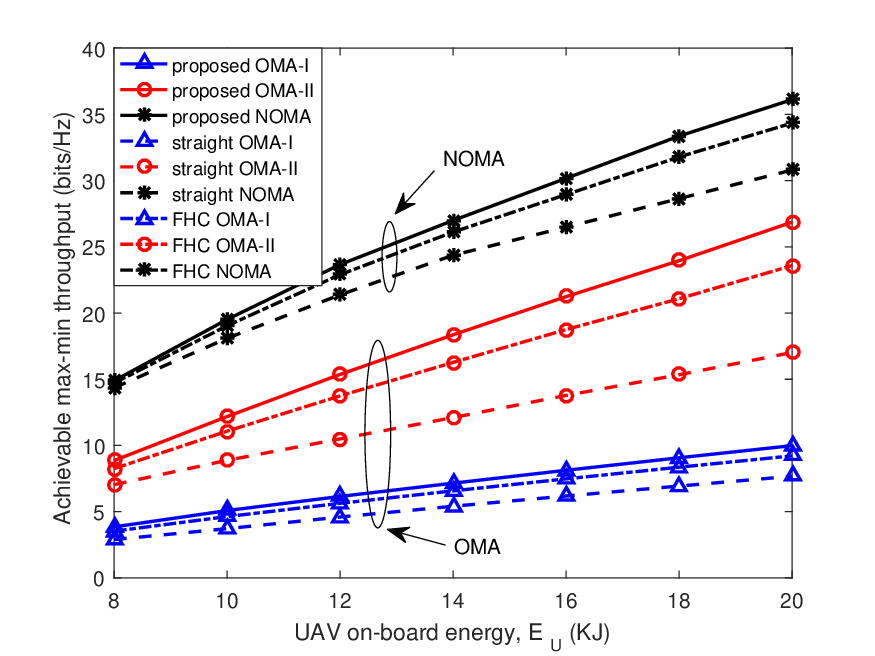}
        \caption{Max-min throughput versus the UAV on-board energy of different schemes for $E_s=10$ J.}
        \label{UAV energy 10-120}
    \end{center}
\end{figure}

In Fig. \ref{UAV energy 10-120}, we provide the max-min throughput versus the UAV on-board energy $E_U$ with $E_s=10$ J for different MA schemes. For comparison, we consider the following benchmark schemes:
\begin{itemize}
  \item \textbf{Straight X}: In this case, the UAV flies from ${\mathbf{u}}_I$ to ${\mathbf{u}}_F$ in a straight line. The corresponding max-min achievable throughput is obtained by solving Problems \eqref{P1.1} and \eqref{P2.1} with additional linear constraints ${\mathbf{u}}\left( {1,n} \right) = {\mathbf{u}}\left( {2,n} \right),\forall n$. Meanwhile, X represents different MA schemes, such as OMA-I, OMA-II, and NOMA.
  \item \textbf{FHC X}: In this case, the UAV collects data from GNs following from the fly-hover-communicate (FHC) protocol as did in \cite{Zeng2019Energy}. The optimization problem becomes to find the optimal hovering locations and the corresponding resource allocations.
\end{itemize}
As illustrated, it is first observed that the max-min throughput of all considered schemes increases with the increase of $E_U$ since the UAV is able to collect more information bits from GNs with a longer flight time. In particular, the performance gain of the proposed scheme or the FHC scheme over the scheme with a straight trajectory becomes more pronounced as $E_U$ increases. This is because a larger value of $E_U$ allows the UAV to fly closer to collect data from GNs, which also validates the benefits of UAV trajectory design. It is also observed that the proposed scheme outperforms the FHC scheme. This is expected since the UAV would consumes more energy to hover in the air, which reduces the total UAV flight time. Moreover, NOMA achieves a better performance than OMA. The performance gain achieved by NOMA comes from the multiplex of all GNs in power domain. Note that OMA-I achieves the worst performance since the UAV needs to allocate time resources equally to all GNs, which reduces the throughput of GNs which have good channel conditions (i.e., near from the UAV).
\begin{figure}[t!]
    \begin{center}
        \includegraphics[width=2.5in]{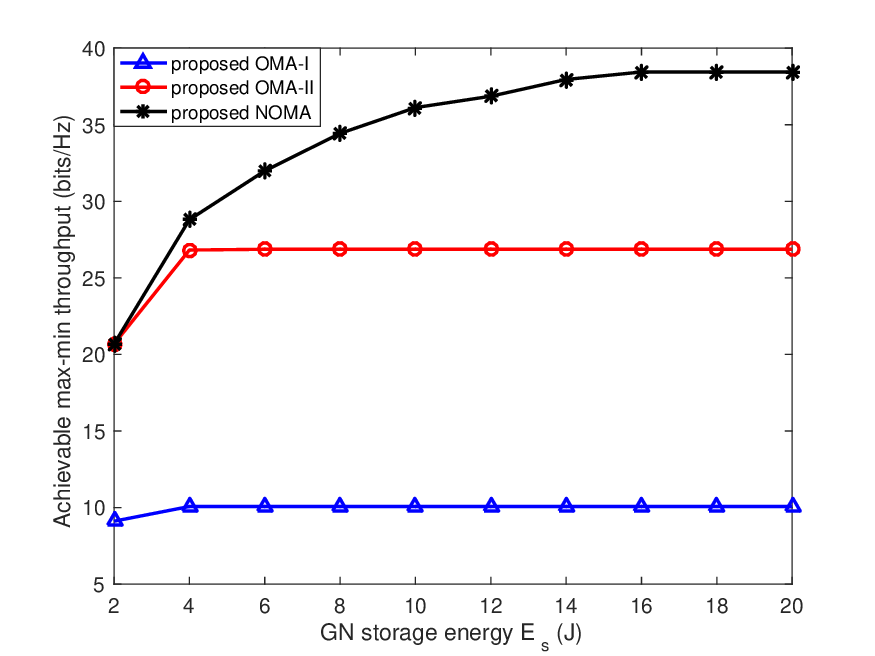}
        \caption{Max-min throughput versus the GN storage energy with $E_U=20$ KJ.}
        \label{Sensor energy 1-20}
    \end{center}
\end{figure}

Fig. \ref{Sensor energy 1-20} presents the max-min throughput versus the GN storage energy $E_s$ for different schemes with $E_U=20$ KJ. From the figure, the max-min throughput of all schemes improves with the increase of $E_s$ at first and remains unchanged. This is because, for a given $E_U$, the UAV can allocate more time to GNs for uploading information when $E_s$ is limited. The max-min throughput remains unchanged until all the UAV flight time is occupied for GNs uploading data with $P_{\max}$. In this case, the increase of $E_s$ has no effect on the obtained max-min throughput since the total energy consumption of GNs is fixed. It is also shown that NOMA always achieves equal or higher max-min throughput than OMA-II. The performance gain of NOMA over OMA becomes more pronounced as $E_s$ increases.
\begin{figure}[t!]
    \begin{center}
        \includegraphics[width=2.5in]{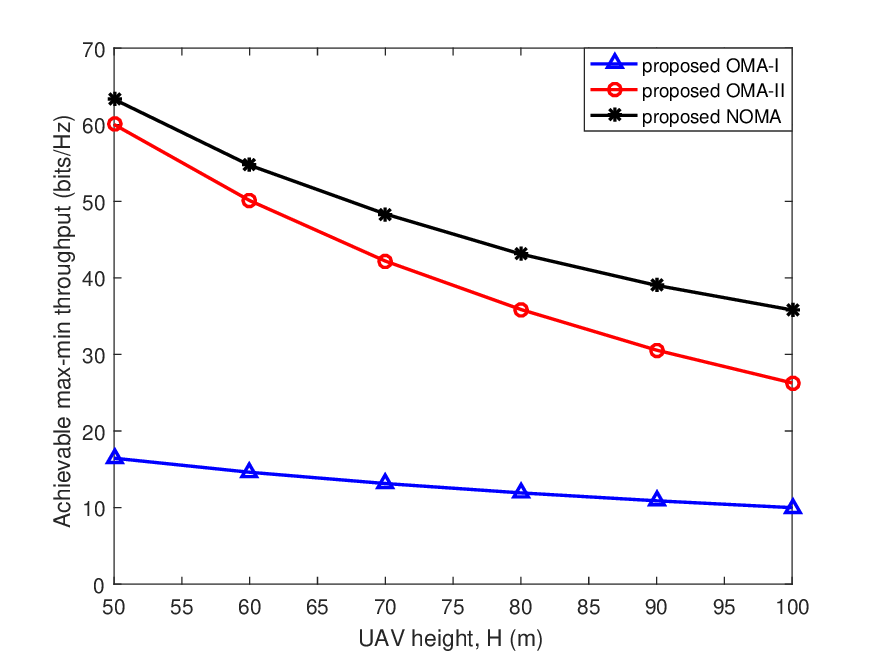}
        \caption{Max-min throughput versus the UAV height for $E_U=20$ KJ and $E_s=10$ J.}
        \label{H 50-100}
    \end{center}
\end{figure}

Finally, Fig. \ref{H 50-100} provides the max-min throughput performances versus the UAV height of different schemes for $E_U=20$ KJ and $E_s=10$ J. It is observed that the performance of all considered schemes decreases with the increase of $H$. This is expected since we assume the deterministic LoS channel model, the A2G channel conditions become weaker when the UAV flies at a higher altitude. It is also observed that the performance degradation of OMA becomes more pronounced than NOMA as $H$ increases, which also confirms the advantages of NOMA transmission scheme.
\section{Conclusions}
In this paper, energy-constrained UAV data collection systems have been investigated with the employment of both OMA and NOMA transmission. The optimization problems for maximization the minimum UAV data collection throughput from all GNs were formulated for the two MA schemes. To solve the resulting non-convex problems, two efficient AO-based algorithms were proposed. For OMA, the original problem was decomposed into two subproblems, which were alternatively solved by applying SCA technique. For NOMA, a penalty-based algorithm was developed to solve the decoding order design subproblem, while other subproblems were solved using SCA technique. Numerical results verified the effectiveness of the proposed designs compared with other benchmark schemes, and demonstrated that the max-min throughput obtained by NOMA is always larger than or equal to OMA.

There are other promising future research directions, some of which are discussed as follows. First, given the considered energy-constrained UAV data collection system, various other performance metrics can be optimized, such as the energy efficiency at both the UAV and GNs. Second, note that only one UAV was assumed in this paper, it is interesting to consider multiple UAVs, especially for the number of GNs is large. In this case, the constraints of collision avoidance among UAVs~\cite{Wu2018} and the management of multi-UAV interference should be considered, which imposes new challenging problems. Moreover, the joint design with other A2G channel models, such as angle-dependent LoS probability model~\cite{Hourani2014Optimal} and Rician fading model, is another interesting but challenging topic in the future work, which may require other sophisticated machine learning tools~\cite{Xiao} to be employed.
\section*{Appendix~A: Proof of Theorem~\ref{penlty}} \label{Appendix:A}
First, the partial Lagrange function of Problem \eqref{P2.7} can be expressed as
\begin{align}\label{partial Lagrange function}
{{\mathcal{L}}}\!\left(\! {{\eta ^{\rm{N}}}\!\!,{\mathbf{A}},{\mathbf{I}},\lambda } \!\right)\! \!=\! {\eta ^{\rm{N}}}\! +\! \lambda \!\left(\! {\sum\nolimits_{k = 1}^K \!{\sum\nolimits_{m \ne k}^K {\!\left(\! {{\alpha _{k,m}}{{\left[ n \right]}^2} \!-\! {\alpha _{k,m}}\left[ n \right]} \!\right)\!} } } \!\right)\!,
\end{align}
where ${\mathbf{A}} = \left\{ {{\alpha _{k,m}}\left[ n \right],\forall k \ne m \in \mathcal{K},n \in \mathcal{N}} \right\}$, ${\mathbf{I}} = \left\{ {{I_k}\left[ n \right],\forall k \in \mathcal{K},n \in \mathcal{N}} \right\}$ and $\lambda $ is the non-negative Lagrange multiplier associated with the constraint \eqref{integer 1}. Therefore, the dual problem of Problem \eqref{P2.7} is
\begin{align}\label{dual problem}
\mathop {\min }\limits_{\lambda  \ge 0} \;\mathop {\max }\limits_{\left( {\eta^{{\rm{N}}} ,{\mathbf{A}},{\mathbf{I}}} \right) \in \mathcal{D}} \;\mathcal{L}\left( {\eta^{{\rm{N}}} ,{\mathbf{A}},{\mathbf{I}},\lambda } \right) = \mathop {\min }\limits_{\lambda  \ge 0} \;\psi \left( \lambda  \right),
\end{align}
where $\mathcal{D}$ is the feasible set spanned by constraints \eqref{decoding order NOMA}, \eqref{Ikn 2.1}, \eqref{sum throughput NOMA P2.6} and \eqref{integer 2} and $\psi \left( \lambda  \right) = \mathop {\max }\limits_{\left( {\eta^{{\rm{N}}} ,{\mathbf{A}},{\mathbf{I}}} \right) \in \mathcal{D}} \;\mathcal{L}\left( {\eta^{{\rm{N}}} ,{\mathbf{A}},{\mathbf{I}},\lambda } \right)$.

Moreover, the primal optimization problem \eqref{P2.7} can be equivalently expressed as
\begin{align}\label{primal problem}
{p^*} = \mathop {\max }\limits_{\left( {\eta^{{\rm{N}}} ,{\mathbf{A}},{\mathbf{I}}} \right) \in \mathcal{D}} \;\mathop {\min }\limits_{\lambda  \ge 0} \mathcal{L}\left( {\eta^{{\rm{N}}} ,{\mathbf{A}},{\mathbf{I}},\lambda } \right).
\end{align}
Due to the weak duality~\cite{convex}, we have the following inequalities:
\begin{align}\label{weak duality}
\begin{gathered}
  \mathop {\min }\limits_{\lambda  \ge 0} \;\psi \left( \lambda  \right) = \mathop {\min }\limits_{\lambda  \ge 0} \;\mathop {\max }\limits_{\left( {\eta^{{\rm{N}}} ,{\mathbf{A}},{\mathbf{I}}} \right) \in \mathcal{D}} \;\mathcal{L}\left( {\eta^{{\rm{N}}} ,{\mathbf{A}},{\mathbf{I}},\lambda } \right) \hfill \\
  \;\;\;\;\;\;\;\;\;\;\;\;\;\;\;\;\; \ge \mathop {\max }\limits_{\left( {\eta^{{\rm{N}}} ,{\mathbf{A}},{\mathbf{I}}} \right) \in \mathcal{D}} \;\mathop {\min }\limits_{\lambda  \ge 0} \mathcal{L}\left( {\eta^{{\rm{N}}} ,{\mathbf{A}},{\mathbf{I}},\lambda } \right) = {p^*}. \hfill \\
\end{gathered}
\end{align}
It is noted that $\sum\nolimits_{k = 1}^K {\sum\nolimits_{m \ne k}^K {\left( {{\alpha _{k,m}}{{\left[ n \right]}^2} - {\alpha _{k,m}}\left[ n \right]} \right)} }  \le 0$ for any ${\mathbf{A}} \in \mathcal{D}$. Thus, $\mathcal{L}\left( {\eta^{{\rm{N}}} ,{\mathbf{A}},{\mathbf{I}},\lambda } \right)$ is a decreasing function with respect to $\lambda $ for ${\left( {\eta^{{\rm{N}}} ,{\mathbf{A}},{\mathbf{I}}} \right) \in \mathcal{D}}$, which means $\psi \left( \lambda  \right)$ is bounded from below by the optimal value of Problem \eqref{P2.7}. Assume that the optimal solutions to the dual problem \eqref{dual problem} are ${{\lambda ^*}}$ and $\left( {{\eta^{N *} },{{\mathbf{A}}^*},{{\mathbf{I}}^*}} \right)$. In the following, we discuss the optimal value of the dual problem \eqref{dual problem} and the equivalent primal problem \eqref{primal problem} in two cases.

First, suppose that $\sum\nolimits_{k = 1}^K {\sum\nolimits_{m \ne k}^K {\left( {{\alpha _{k,m}^*}{{\left[ n \right]}^2} - {\alpha _{k,m}^*}\left[ n \right]} \right)} }  = 0$. Since $\left( {{\eta ^{N *}},{{\mathbf{A}}^*},{{\mathbf{I}}^*}} \right)$ are also feasible to Problem \eqref{primal problem}, we have the following inequalities:
\begin{align}\label{inequalities 2}
{p^*} \ge {\eta ^{N *}} = \mathcal{L}\left( {{\eta ^{N *}},{{\mathbf{A}}^*},{{\mathbf{I}}^*},{\lambda ^*}} \right) = \psi \left( {{\lambda ^*}} \right).
\end{align}
Based on \eqref{weak duality} and \eqref{inequalities 2}, we have
\begin{align}\label{zero gap}
\begin{gathered}
  \mathop {\max }\limits_{\left( {{\eta ^{{\rm{N}}}},{\mathbf{A}},{\mathbf{I}}} \right) \in {{\mathcal{D}}}} \;\mathop {\min }\limits_{\lambda  \ge 0} {{\mathcal{L}}}\left( {{\eta ^{{\rm{N}}}},{\mathbf{A}},{\mathbf{I}},\lambda } \right) \hfill \\
  \;\;\;\;\;\;\;\;\;\;\;\;\;\;\;\;\; = \mathop {\min }\limits_{\lambda  \ge 0} \;\mathop {\max }\limits_{\left( {{\eta ^{{\rm{N}}}},{\mathbf{A}},{\mathbf{I}}} \right) \in {{\mathcal{D}}}} \;{{\mathcal{L}}}\left( {{\eta ^{{\rm{N}}}},{\mathbf{A}},{\mathbf{I}},\lambda } \right), \hfill \\
\end{gathered}
\end{align}
which implies the strong duality between the equivalent primal problem \eqref{primal problem} and the dual problem \eqref{dual problem} holds when $\sum\nolimits_{k = 1}^K {\sum\nolimits_{m \ne k}^K {\left( {{\alpha _{k,m}}{{\left[ n \right]}^2} - {\alpha _{k,m}}\left[ n \right]} \right)} }  = 0$. Recall from that $\mathcal{L}\left( {\eta^{{\rm{N}}} ,{\mathbf{A}},{\mathbf{I}},\lambda } \right)$ is a decreasing function with respect to $\lambda $, we have
\begin{align}\label{decrease}
\psi \left( \lambda  \right) = {p^*},\forall \lambda  \ge {\lambda ^*}.
\end{align}

Second, when $\sum\nolimits_{k = 1}^K {\sum\nolimits_{m \ne k}^K {\left( {{\alpha _{k,m}^*}{{\left[ n \right]}^2} - {\alpha _{k,m}^*}\left[ n \right]} \right)} }  < 0$, $\psi \left( {{\lambda ^*}} \right) = \mathop {\min }\limits_{\lambda  \ge 0} \;\psi \left( \lambda  \right) \to  - \infty $ due to the monotone decreasing of $\psi \left( \lambda  \right)$ with respect to $\lambda $. This contradicts the inequality in \eqref{weak duality} since ${p^*}$ is a finite value.

Therefore, $\sum\nolimits_{k = 1}^K {\sum\nolimits_{m \ne k}^K {\left( {{\alpha _{k,m}^*}{{\left[ n \right]}^2} - {\alpha _{k,m}^*}\left[ n \right]} \right)} }  = 0$ must hold at the optimal solution and the proof of Theorem~\ref{penlty} is completed.
\bibliographystyle{IEEEtran}
\bibliography{mybib}

\end{document}